\documentclass[a4paper,onecolumn,11pt,accepted=2024-12-15]{quantumarticle}
\pdfoutput=1
\usepackage{graphicx}
\usepackage{amsfonts}
\usepackage{amsmath}
\usepackage[utf8]{inputenc}
\usepackage[english]{babel}
\usepackage[T1]{fontenc}
\usepackage{amsmath}
\usepackage{hyperref}
\usepackage[numbers,sort&compress]{natbib}
\usepackage{tikz}
\usepackage{lipsum}

\def\ee{\end{equation}}
\def\bea{\begin{eqnarray}}

\def\bra#1{\langle #1 |}
\def\ket#1{| #1\rangle}

\def\braket#1#2{\langle \, #1 \, | \, #2 \, \rangle}

\def\Tr{{\rm Tr}}

\begin{document}

\title{The measurement postulates of quantum
mechanics are not redundant}

\author{Adrian Kent}
\email{apak@cam.ac.uk} 
\affiliation{Centre for Quantum Information and Foundations, DAMTP, Centre for
  Mathematical Sciences, University of Cambridge, Wilberforce Road,
  Cambridge, CB3 0WA, U.K.}

\affiliation{Perimeter Institute for Theoretical Physics, 31 Caroline Street North, Waterloo, ON N2L 2Y5, Canada.}

\maketitle

\begin{abstract}
  Masanes, Galley and M\"uller \cite{masanes2019measurement} argue that the measurement
  postulates of non-relativistic quantum mechanics follow from the structural postulates
  together with an assumption they call the ``possibility of state
  estimation''.   Their argument also relies on what they term a ``theory-independent
  characterization of measurements for single and multipartite
  systems''.   We refute their conclusion, giving explicit examples of
  non-quantum measurement and state update rules that satisfy all their assumptions.
  We also show that their ``possibility of state estimation'' assumption is neither necessary nor sufficient to ensure
  a sensible notion of state estimation within a theory whose states
  are described by the quantum formalism. 
  We further show their purportedly ``theory-independent'' characterization assumes several properties of
  quantum measurements that exclude plausible alternative
  types of measurement.  We illustrate all these points with specific alternative
  measurement postulates and post-measurement state update rules.

  We conclude that, contrary to some folklore, quantum
  mechanics is by no means an island in theory-space.  It 
  can consistently be extended by rules for obtaining
  information about quantum states other than via POVMs.   Whether
  such rules are realised in nature, for example in linking quantum
  theory and gravity, is an empirical question that cannot be
  resolved by theoretical analysis alone. 

\end{abstract}

\section{Introduction}

No-go theorems ruling out some types of extension or
alternative to quantum mechanics have played a crucial role
in advancing our understanding of fundamental physics. 
In particular, they help delineate the scope for new solutions to two of the deepest problems in contemporary physics, the
quantum measurement problem and the unification of 
quantum theory and gravity.

Alas, ``zombie theorems'' have also propagated.
Von Neumann famously argued that ``an introduction of
hidden parameters is certainly not possible without a
basic change in the present [quantum] theory'', and
claimed to show that ``the present system of quantum
mechanics would have to be objectively false, in order
that another description of the elementary processes
than the statistical one be possible'' (pp. 210 and 325 of \cite{von2018mathematical}).
As Bell pointed out \cite{bell1966problem}, von Neumann's argument
fails, 
and indeedd his conclusion was explicitly refuted by Bohm \cite{bohm1952suggested},
following the earlier  ideas of de Broglie \cite{debroglie1927}.  
Von Neumann assumed that the linearity of expectation values
of a combination of observables that holds in quantum theory
should also hold for states defined by hidden variables,
which produce deterministic outcomes to the measurement of
any observable.  This does indeed exclude the possibility of
such states, but it is quite unreasonable.
Yet, Bell noted \cite{bell1982impossible}, false impossibility ``proofs''
continued to be proposed at least as late as 1978 \cite{jost1978comment}.  

Another example is Eppley and Hannah's claim \cite{eppley1977necessity} 
that a contradiction would
arise in any dynamical theory in which a classical gravitational field
interacts with quantum matter.
In their words
\begin{quote}
$\ldots$ we show that if a gravitational wave of arbitrarily small 
momentum can be used to make a position measurement on a quantum particle,
i.e., to ``collapse the wave function into an eigenstate of
position,'' then the
uncertainty principle is violated. If the interaction does not result in collapse
of the wave function, it is then possible to distinguish experimentally between
superposition states and eigenstates. We show that this ability allows one to
send observable signals faster than $c$ when applied to a state consisting of two
spatially separated particles with correlated spins.
\end{quote}
This has been challenged on a variety of grounds \cite{huggett2001quantize,mattingly2006eppley},
among them that Eppley-Hannah's discussion of quantum measurements on entangled
systems is simply incorrect \cite{albers2008measurement,kent2018simple}.  
In fact, a simple model refutes Eppley-Hannah's claim that superluminal
signalling necessarily follows from their assumption \cite{kent2018simple}.

Confirmation bias seems to be a factor in these and other fallacious
arguments and misrepresentations.    
Many theorists believe that quantum theory is so elegantly and
delicately coherent that any alteration must necessarily be
inconsistent.   Many also believe
there is no alternative to a quantum theory of gravity.
Purported proofs of these hunches are often uncritically welcomed, since they
justify doing what many theorists want to do anyway, which is
not only to focus on quantum theory and quantum gravity
but also to dismiss alternatives and concerns without
further discussion.  
The former is a perfectly defensible theoretical choice, but the
latter is not. We should be alert to our biases and learn from the history of misclaims.
No-go theorems purporting to exclude large classes
of non-quantum or post-quantum theories need careful analysis. 

This paper looks at interesting recent work \cite{masanes2019measurement} by Masanes, Galley and
M\"uller (MGM) on the logical relationship between the postulates of
quantum mechanics.  MGM claim that the measurement postulates of
quantum mechanics are ``operationally redundant''.
This claim, in the title and elsewhere, appears to suggest that the
measurement postulates can be derived
from the other postulates of quantum mechanics.
In fact, though, even MGM recognize this is not true.   
They appeal to an additional strong assumption, which they call ``the possibility of
state estimation''.   However, this terminology too is misleading, as
we will explain, in that it neither characterizes the possibility of state estimation for,
nor gives a natural constraint on, extensions of quantum
mechanics. 

The qualification {\it operationally} redundant also needs careful discussion.
At first reading, one might perhaps take it to mean something like
``redundant if we treat preparation and measurement devices
as black boxes with inputs and outputs'' or ``redundant within
the Copenhagen framework'' or even ``redundant within any
sensible view of quantum mechanics''.    
MGM say they
\begin{quote}
  ``$\ldots$ take an operational approach, with the notions
of measurement and outcome probability being primitive elements
of the theory, but without imposing any particular structure
on them.''
\end{quote}
In fact, though, their approach
involves many assumptions about the form and properties
of measurements, including some that are not explicitly stated.
Several of these are not natural constraints on extensions of quantum
mechanics, from an operational perspective or otherwise.

Even allowing for all their assumptions, MGM's claim to derive
the quantum measurement postulates is incorrect.
We refute it by giving alternative 
measurement postulates that have all the properties MGM require
but that do not satisfy the quantum measurement postulate or
state-update rule.   The key insight underlying our alternative postulates is
that quantum mechanics can consistently be extended by hypothetical
measurement devices
that give complete information about the state of a subsystem \cite{kent2005nonlinearity}.
It is consistent to postulate that these hypothetical measurement devices 
leave the state unaltered; it is also consistent to postulate that
they alter the state via a map that depends on the measurement outcome,
which leads to consistent non-linear versions of quantum theory.
These results were first proven in Ref. \cite{kent2005nonlinearity}.  Further
examples and applications of measurement devices that give partial
and/or stochastic information were given in
Ref. \cite{kent2021quantum}.  
These examples were given in the context of relativistic
quantum mechanics. {\it A fortiori}, they also define consistent
alternatives to, or extensions of, non-relativistic
quantum mechanics, which is the focus of MGM's discussion.

We give several more examples of theoretically interesting hypothetical
devices here.  These give partial information about the quantum state
of a subsystem.  They can thus be constructed from devices that give
complete information about the quantum state, together with suitable
classical post-processing devices and randomness generators.
They are of potentially independent interest, since they suggest
other ways in which nature might allow
measurements that go beyond quantum measurements, some
of which might possibly arise in theories that combine
quantum theory with classical and/or other degrees of freedom.

We show that some of these devices define alternative measurement
postulates that refute MGM's purported derivations of the quantum
measurement postulates.   
We also discuss the implications of others for MGM's approach.
These latter devices define measurement postulates that do not satisfy all of MGM's assumptions
and so do not provide further direct counterexamples to MGM's
purported derivation.   However, they do define further interesting
and consistent extensions of quantum mechanics, which MGM
suggest -- in the title and elsewhere in Ref. \cite{masanes2019measurement} --
should not exist.   They highlight (i) some
tacit assumptions about possible measurement rules that MGM make and
that seem hard to justify, (ii) that some of their explicit assumptions
are significantly and unnecessarily restrictive, and (iii) 
weaknesses in some of their justifications.  
In particular, they highlight the problems mentioned above with MGM's (so-called) ``possibility of
state estimation'' assumption: namely, that it is neither necessary nor
sufficient to establish the possibility of state estimation.

In short, they highlight that MGM's definition of a measurement postulate
excludes alternative measurement postulates that seem natural, interesting
and even potentially physically relevant.
This suggests that it is unlikely to be possible to justify any definition
similar to MGM's that would rescue their intended results.

The measurement postulates
of quantum theory are not redundant, but an essential part of the theory's definition.
Whether they describe the only way to obtain information from
quantum states is an open empirical question.
We should keep an open mind and investigate it in untested regimes, for example where
delocalized mesoscopic masses have measurable gravitational effects
(see e.g. \cite{bose2017spin,marletto2017gravitationally,van2020quantum,howl2021non,PRXQuantum.2.030330,kent2021testing,kent2021testing2}).

\section{MGM's characterisation of quantum mechanics}

\subsection{Assumptions MGM do not make} 

We first note assumptions that might be thought necessary
(if not necessarily sufficient) to characterise quantum
mechanics but that MGM do not make.

\subsubsection{\bf MGM do not assume no-signalling}

MGM consider only non-relativistic quantum mechanics.
In particular, the relativistic no-signalling principle
plays no role in their discussion.   Nor does the
quantum no-signalling principle, which ensures that
standard quantum measurements on separated entangled systems
do not give a signalling mechanism.
In fact, MGM assume nothing at all about measurements on
an entangled subsystem.  
We discuss these issues further below, following MGM in focussing
on non-relativistic quantum mechanics.   

\subsubsection{\bf MGM do not assume measurements are repeatable}

Some results characterising quantum measurements assume
some form of repeatability, for example that successive measurements
should have the same outcome when there is no
intervening evolution.   This holds
for projective measurements but not for general
POVMs.  MGM's measurement postulates cover
general quantum measurements, and so they
do not assume this version of repeatability.

In fact, they make no explicit postulate about successive
measurements.  In particular, they do not postulate that
two or more measurements applied in sequence to a
quantum system can be considered as a single measurement
on the system.

\subsection{MGM's postulates}

MGM characterise non-relativistic quantum mechanics by three structural
postulates together with two measurement postulates.
\newline

\textbf{Postulate (states).} To every physical system there
corresponds a complex and separable Hilbert space $ \mathbb{C}^d $,
and the pure states of the system are the rays of $P
\mathbb{C}^d $.

The rays are equivalence classes under non-zero complex
multiplication.  Following MGM we represent these by normalised states 
$ \psi \in \mathbb{C}^d $, and use the notation $\mathbb{C}^d $ to
represent finite-dimensional Hilbert spaces and countably 
infinite-dimensional Hilbert spaces (denoted by $d= \infty$).
\newline

\textbf{Postulate (transformations).} The reversible transformations
(for example, possible time evolutions) of pure states of
$\mathbb{C}^d$
are the unitary transformations $ \psi \rightarrow U\psi $ with $ U
\in U(d) $.
\newline

\textbf{Postulate (composite systems).} The joint pure states of
systems $\mathbb{C}^a$ and $\mathbb{C}^b$ are the rays of the tensor product Hilbert space $\mathbb{C}^a \otimes \mathbb{C}^b$.
\newline

\textbf{Postulate (measurement).} Each measurement outcome of system $\mathbb{C}^d$ is represented by a linear operator $Q$ on $\mathbb{C}^d$ satisfying $0 \leq Q \leq I$, where $I$ is the identity. The probability of outcome $Q$ on state $\psi \in \mathbb{C}^d$ is given by
\begin{equation}
  P(Q|\psi) = \langle \psi | Q | \psi \rangle \, .
\end{equation}

A (full) measurement is represented by the operators corresponding to
its outcomes $Q_1,\ldots,Q_n$, which must satisfy the normalization
condition
\begin{equation}
  \sum_{i=1}^n Q_i = I \, . 
\end{equation}

MGM do not state whether $n$ may be infinite.   We will return to this
point later.  
\newline

\textbf{Postulate (post-measurement state-update).} Each outcome is represented by a completely-positive linear map $\Lambda$ related to the operator $Q$ via
\begin{equation}
\mathrm{tr}(\Lambda(|\psi \rangle \langle \psi|)) = \langle \psi | Q |
\psi \rangle, \qquad \text{for all } \psi  \, . 
\end{equation}
The post-measurement state after outcome $\Lambda$ is
\begin{equation}
\rho = \frac{\Lambda(|\psi \rangle \langle
  \psi|)}{\mathrm{tr}(\Lambda(|\psi \rangle \langle \psi|))} \, . 
\end{equation}
A (full) measurement is represented by the maps corresponding to its outcomes $\Lambda_1, \ldots, \Lambda_n$ whose sum
\begin{equation}
\sum_{i=1}^n \Lambda_i
\end{equation}
is trace-preserving.
\newline

After presenting these postulates, MGM state
they will
\begin{quote}
  ``prove that the “measurement” and “post-measurement
state-update” postulates are a consequence of the
first three postulates.''  (p. 2 of Ref. \cite{masanes2019measurement})
\end{quote}
Note that neither the term ``operationally'' nor any other
qualification are mentioned at this point.
This illustrates how hard it can be to avoid unintentionally
overstating limited technical no-go results.
It is important to be clear that MGM actually argue that the measurement and state-update
postulates follow from the three structural postulates {\it together
  with other strong assumptions}, which we review below.   In other words, MGM do not
even try to prove what the quote above claims.  

\section{MGM's characterisation of a measurement postulate}

MGM give what they describe as a ``[f]ormalism for any alternative
measurement postulate'' baseed on ``a theory-independent
characterization of measurements for single and multipartite
systems'', on pp. 3-4 of Ref. \cite{masanes2019measurement}.
Earlier, on p.2, they make some comments
on the structure of mixed states in quantum theory and
hypothetical alternative theories.   We address these
first, since they are important for our discussion.

\subsection{MGM and the role of mixed states}

\subsubsection{\bf Proper mixed states}

MGM use ``mixed states'' to refer to what are
generally called ``proper mixed states'', that is,
probabilistic mixtures of pure states.
More precisely, MGM say a mixed state
\begin{quote}
``is an equivalence class of indistinguishable ensembles,
and an ensemble $( \psi_r , p_r )$  is a probability distribution over pure
states.''
\end{quote}
Their notation is ambiguous, but we will generally assume it includes
both finite and countably infinite ensembles.
Where we restrict to finite ensembles, or allow
uncountably infinite ensembles that have a well-defined quantum
density matrix, we will state this explicitly. 
It will not matter for our counterexamples to MGM's purported
derivation of the quantum measurement postulates whether the ensembles are
restricted to be finite or countably infinite.

MGM note that mixed states are not mentioned in
the standard postulates of quantum mechanics.    
Indeed, a mixed state is
not normally considered to be a fundamental
notion in quantum mechanics.  We normally consider
ensembles in contexts where our knowledge
of the true state is imperfect: some other
agent, or nature, has prepared a specific pure state $\psi_r$ 
drawn from the ensemble, but we know only the
ensemble $( \psi_r , p_r )$.

In principle, the postulates of quantum mechanics
can be applied directly to ensembles without referring
to density matrices: if $( \psi_r , p_r)$
undergoes a time evolution $U$, the resulting ensemble
is $(U \psi_r , p_r )$, and so on.
In fact, of course, quantum mechanics allows a much simpler treatment,
in which the ensemble $( \psi_r , p_r )$ is represented by
the density matrix
\begin{equation}\label{dmdef}
\rho = \sum_r p_r \ket{ \psi_r } \bra{ \psi_r } \, .
\end{equation}

A time evolution $U$ produces
\begin{equation}
  U \rho U^{\dagger}
\end{equation}

and a measurement with outcome associated to the operator $Q$ and completely positive linear map
$\Lambda$ has probability
\begin{equation}
 \Tr ( Q \rho )
\end{equation}
and post-measurement density matrix
\begin{equation}
  \rho' = \frac{ \Lambda ( \rho )  }{  \Tr ( \Lambda ( \rho ) )  } \, .
\end{equation}

Because the density matrix $\rho$ contains all the information about
the possible outcomes and probabilities of any sequence of
measurements on the ensemble, two ensembles with the same density
matrix are indistinguishable, and hence belong to the same
equivalence class.   Hence in quantum mechanics the
equivalence class of the ensemble $( \psi_r , p_r )$ is represented by
the density matrix (\ref{dmdef}).

As MGM stress, this follows from the measurement postulates of
quantum mechanics.  As they note, different
measurement postulates could imply different equivalence relations and
hence different equivalence classes of ensembles and a different set
of mixed states.
MGM thus do not assume that proper mixed states
must necessarily be represented in the form (\ref{dmdef}). 

We would argue that in fact one need not make any assumption about
measurements on proper mixed states, since measurement postulates (quantum or otherwise) can be applied
directly to ensembles.  The action
of a measurement on an ensemble is determined by its action on the
pure states in the ensemble, and so we only need postulates that determine
the outcome and effect of measurements on pure states.
For the same reason, an open-minded
analysis of possible non-quantum measurement postulates need not --
indeed, we
would argue, should not -- assume there must be any non-trivial equivalence
relation between ensembles.  An alternative to or extension of quantum
theory can still be well-defined and interesting even if it implies
no ensemble equivalences.
After all, MGM aim (and claim) 
to show that no alternative measurement postulates exist.
Whether some alternative measurement postulates require less convenient calculations than quantum theory
is a separate (and secondary) question.

We would also argue that, whereas a 
pure state is a fundamental notion in quantum theory, an ensemble is
not.  On the standard view of quantum theory underlying any ensemble
description of a physical system there always is an underlying pure
state of a suitably chosen larger system (for example, the universe,
although smaller systems may suffice). \footnote{We do not mean to
  exclude the possibility that there is not (or not always) a pure state
  of the universe.    For example, the most complete cosmological theory possible
  might use a density matrix to define
  the initial conditions.  But an initial state density matrix is not
  associated with a specific ensemble, so MGM's discussion of
  ensembles would not directly apply.  In particular their ``possibility of state
  estimation'' assumption would have different implications if it
  applied to density matrices rather than ensembles.  
  In any case, this possibility is not discussed by MGM, so we do not
  pursue it here. }

MGM take a different view of the status of proper mixed states defined
by ensembles.   These play a crucial role in their
assumption of the ``possibility of state estimation'', 
which is meant
to hold for ensembles as well as pure states.
This postulate requires that ensembles of states
in $\mathbb{C}^d$ should be characterized by the outcome
probabilities of some finite set of measurements.
Since infinitely many parameters are needed to characterise
the set of ensembles of states, this implies that there are equivalence
classes whose ensembles require infinitely many parameters to characterise.
As we discuss below, that excludes otherwise interesting and consistent alternative
measurement postulates

\subsubsection{\bf Improper mixed states}

MGM's discussion of mixed states does not apply to the so-called
``improper mixed states'' that arise from entanglement with other
systems.\footnote{The terminology is confusing.
  So-called improper mixed states are fundamental in quantum theory,
  while so-called proper mixed states are either improper mixed
  states in a suitably enlarged Hilbert space or secondardy quantities
  defined by combining quantum theory and classical probabilities.
  However, the terminology has unfortunately become standard, so we
  retain it here.}

This is a major gap in their analysis. Their states and composite systems postulates imply
that the states of $n$ systems are represented by
normalised states $\psi \in \mathbb{C}^{d_1} \otimes \ldots \otimes
\mathbb{C}^{d_n}$.   Their transformations postulate implies that,
even if the state is initially a product state, it can (and
generically will) evolve into an entangled state.
In particular, the reduced density matrix
\begin{equation}\label{reddmone}
\rho_1 = \Tr_{d_2 \ldots d_n} ( \ket{\psi} \bra{ \psi} ) \in O (
\mathbb{C}^{d_1} ) \, 
\end{equation} of
system $1$ may (and generically will) not represent a pure
state in  $\mathbb{C}^{d_1}$.
So, pace MGM, a discussion of quantum measurement postulates, or alternative
measurement postulates in the quantum framework, necessarily has
to consider measurements on improper mixed states.

The alternative measurement postulates we set out below are explicitly
defined on improper
mixed states, mostly via the reduced density matrix (\ref{reddmone}).

\subsubsection{\bf Distinguishing proper and improper mixtures}

At first sight, the distinction between proper and improper mixed
states might seem hard to maintain.
If \mbox{$\psi \in \mathbb{C}^{d_1} \otimes \ldots \otimes
\mathbb{C}^{d_n}$} is entangled then, from the perspective of
an observer who only has access to system 1, measurements on
systems $2, \ldots , n$ generally replace $\psi$ by an
ensemble of states corresponding to the possible measurement
outcomes, with the corresponding outcome probabilities.
If $\psi$ is an entangled state of the first system with the rest, it
has Schmidt decomposition
\begin{equation}\label{schmidt}
\psi = \sum_{i=1}^k ( p_i )^{\frac{1}{2}} \phi_i \otimes \chi_i  \, ,
\end{equation}
where $k>1$, the $p_i >0$, $\sum_{i=1}^k p_i =1 $,
the states $\phi_i \in {\mathbb C}^{d_1}$ are
orthonormal, and the states
$\chi_i \in \mathbb{C}^{d_2} \otimes \ldots \otimes
\mathbb{C}^{d_n}$ are orthonormal.
A standard quantum projective measurement onto a
basis of $\mathbb{C}^{d_2} \otimes \ldots \otimes
\mathbb{C}^{d_n}$ that includes the $\chi_i$ then
results in outcome $i$ and the post-measurement state
\begin{equation}
\psi'_i = \phi_i \otimes \chi_i
\end{equation}
with probability $p_i$.

For an observer who has access only to system $1$
and is ignorant of the measurement outcome, the state
of system $1$ before the measurement is described by the reduced
density matrix (\ref{reddmone}) representing the improper mixed state, i.e., by
\begin{equation}\label{postmdm}
\rho_1 = \Tr_{d_2 \ldots d_n} ( \ket{\psi} \bra{ \psi} ) = \sum_{i=1}^k p_i \ket{\phi_i } \bra{\phi_i} \, .
\end{equation}
For the same observer, the state of system $1$ after the measurement is described
by the ensemble $ ( \phi_i , p_i ) $, a proper mixed state
also represented by the density matrix on the right hand side of
Eqn. (\ref{postmdm}).
This locally imperceptible transition between improper and proper
might suggest they should be treated as equivalent when considering
possible measurement postulates. 

In fact, though, the true post-measurement state
of the $n$ systems is a pure state of the form
\begin{equation} \label{postmeasstate}
  \phi_i \otimes \chi_i \, ,
\end{equation}
where $i$ is the measurement outcome. 
No physical principle in non-relativistic quantum mechanics precludes an observer
accessing system $1$ from being immediately aware of the outcome
of measurements on the other systems: information can
be communicated instantaneously in non-relativistic
quantum mechanics.   So we can consistently maintain the
distinction between the pre-measurement improper mixed
state given by the reduced density matrix (\ref{reddmone})
and the post-measurement state.
It is only for observers who lack available information that the
latter is described by an ensemble whose proper mixed state has density matrix
given by the right hand side of Eqn. (\ref{postmdm}).  

Although this goes beyond the scope of MGM's discussion, it is
important to emphasize
this distinction can also consistently be maintained in relativistic
quantum mechanics when the systems are spatially separated, by defining the reduced state 
so as to allow for quantum measurements within, but not
outside, the past light cone \cite{kent2005nonlinearity}.

In summary, quantum or alternative measurement postulates need to
describe measurements on improper mixed states.
They do not, we have argued, also need to describe measurements on proper
mixed states.   However, MGM's postulates apply to proper
mixed states, which makes them ill-motivated.
They do not apply to improper mixed states, which makes MGM's postulates
under-defined and, as we will see, also allows a simple refutation of MGM's
claimed derivation of the quantum measurement postulates.   

\subsection{MGM's notion of outcome probability function}

\subsubsection{\bf OPFs and Contextuality}

MGM state (p.3 of Ref. \cite{masanes2019measurement}) that their ``theory-independent characterization of
measurements $\ldots$ is based on the concept of outcome
probability function (OPF)''.   According to their definition, 
each measurement outcome that can be observed on
system $\mathbb{C}^d$ is represented by the
function $f: P \mathbb{C}^d \rightarrow [0,1]$
that is defined as the probability
$f  ( \psi ) = P( f | \psi )$ for
each pure state $\psi \in P \mathbb{ C}^d$.
The complete set of OPFs of system $\mathbb{C}^d$
is denoted by ${\mathcal F}_d$.  
A full measurement with $n$ outcomes is specified
by the OPFs $f_1 , \ldots , f_n$ corresponding to
each outcome, which must satisfy
\begin{equation}
  \label{fullmment}
\sum_{i=1}^n f_i ( \psi ) = 1 \qquad{\rm for~all~pure~states~}\psi \,
.
\end{equation}

MGM stress (p. 4) that this definition is not intended to
imply that, if outcomes are associated to elements of bases
or more generally to positive operators, the outcome probabilities
are necessarily independent of the chosen basis or decomposition
of the identity.   That is, they do not assume non-contextuality.
A measurement outcome thus needs to be understood as defined
relative to a specified full measurement, and its probability function
generally depends on that full measurement.
So, for example, the basis selection devices we describe below can be included
within the OPF framework, although the probability that
the outcome is a given state depends on which basis including
the state is chosen.

\subsubsection{\bf Properties of OPFs}

MGM define a complete set ${\mathcal F}_d$ of OPFs
of system $\mathbb{C}^d$ to be one that is closed
under taking mixtures, composition
with unitaries, and systems composition.
These are defined as follows:
\newline
\newline
\textbf{Property 1 (${\mathcal F}_d$  is closed under taking mixtures):}
Suppose that the random variable $x$ with probability $p_x$ determines
which 2-outcome measurement $\{f_1^x, f_2^x \} \in F_d$ we
implement, and later on we forget the value of $x$.
Then the probability of outcome $1$ for this ``averaged'' measurement is
\begin{equation}\label{opfmix}
    \sum_x p_x f_1^x \in {\mathcal F}_d \, , 
\end{equation}
which must be a valid OPF. Therefore, mixtures of OPFs are OPFs.
\newline
\newline
\textbf{Property 2 (${\mathcal F}_d$ is closed under composition with
  unitaries):}
We can always perform a transformation $U \in U(d)$ before a
measurement $f \in {\mathcal F}_d$, effectively implementing the measurement
\begin{equation}
    f \circ U \in  {\mathcal F}_d \, , 
\end{equation}
which then must be a valid OPF.
\newline
\newline
\textbf{Property 3 ((${\mathcal F}_d$ is closed under systems
  composition):} Since ${\mathcal F}_d$ is complete, it also includes
the measurements that appear in the description of $\mathbb{C}_d$ as part of
the larger system $\mathbb{C}_d \otimes \mathbb{C}_b \simeq \mathbb{C}_{db}$, for any background
system $\mathbb{C}_b$.
Formally, for each background state $\phi \in \mathbb{C}_b$ and global
OPF $g \in {\mathcal F}_{db}$ there is a
local OPF $f_{\phi,g} \in {\mathcal F}_d$ which represents the same
measurement outcome: 
\begin{equation}
    f_{\phi,g}(\psi) = g(\psi \otimes \phi) 
  \end{equation}
  for all $\psi \in P \mathbb{C}_d$.
\newline
\newline

MGM motivate Eqn. (\ref{opfmix}), and
hence Property $1$, only for $2$-outcome
measurements.
They do not postulate that if $f$ is a measurement outcome then
there is necessarily an outcome $I-f$, represented by the
function $(1-f)$ that gives the probability $ (1-f)(\psi) =
1 - P(f | \psi )$ for each $\psi \in P \mathbb{C}^d$, though
this would seem natural.
Nor do they postulate that for each $n$-outcome measurement $M= \{ f_1 , \ldots
, f_n \}$, with $n>2$, and each nontrivial partition  $S_1 \cup S_2 = \{ 1 , \ldots
, n \}$ of the outcomes, 
there is a $2$-outcome measurement $M' = \{ g_1 , g_2 \}$,
where $g_i = \sum_{j \in S_i} f_j$.   This would also seem
natural, since $M'$ can be implemented by applying $M$ and
then forgetting all information about the outcome except the
$S_i$ to which it belongs.

Without some such assumptions, the motivation given by MGM for
Property 1 implies no constraint on $n$-outcome measurements
for $n>2$.  However, we will take Property $1$ to require
that any mixture (i.e. convex combination) of OPFs is an OPF, 
as this is stated in the text, albeit not
fully motivated there.

\subsection{MGM's ``possibility of state estimation'' assumption}

MGM's arguments rely crucially on
a further assumption, which they describe as follows:
\newline

{\it \bf Assumption (possibility of state estimation).} Each finite-dimensional
system ${\mathbb C}^d$ has a finite list of outcomes $f^1 , \ldots , f^k \in
{\mathbb F}_d $ such that knowing their value on any ensemble
$( \psi_r , p_r )$
allows
us to determine the value of any other OPF $g \in
{\mathbb F}_d $ 
on the
ensemble $( \psi_r , p_r )$.

\section{Setting for our post-quantum measurement postulates}

We define here some specific examples of possible types of measurement
that are not allowed within standard quantum mechanics.
For the purposes of our discussion, we consider these as
defining a ``post-quantum'' mechanics, in which one or more type of
non-quantum measurement is allowed as well as all standard
quantum measurements.   Quantum measurements follow the measurement
and post-measurement state-update rules given above; non-quantum
measurements need not necessarily follow either rule.   

It is also interesting to consider the scope for more
general theories in which the structural quantum postulates hold
but the quantum measurement postulates do not.
Howver, we do not consider such theories here, 
focusing on how specific
alternative measurement postulates challenge MGM's assumptions and arguments.   

Following MGM, we consider quantum mechanics for a set of $n$ physical systems,
where $1 \leq n \leq \infty$: here and below we use $\infty$ to denote countable
infinity.  These systems are labelled by $i$, where $1 \leq i \leq
n$, and are represented by rays in complex vector spaces of dimension
$d_i$, where $2 \leq d_i \leq \infty$.  The composite systems postulate tells us that the states
are rays in  $\mathbb{C}^{d_1} \otimes \ldots \otimes
\mathbb{C}^{d_n}$.  (We abuse this notation to mean
$\mathbb{C}^{d_1} \otimes
\mathbb{C}^{d_2} \otimes \ldots$ in the case $n = \infty$.)

MGM do not discuss the relation of the systems considered to the
rest of the universe, but it will be important for our discussion.
We take the product space to be large enough that
there is no entanglement with
any other systems, so that the state of the $n$ systems is indeed a
ray in $\mathbb{C}^{d_1} \otimes \ldots \otimes
\mathbb{C}^{d_n}$.   Depending on the context and the
version of quantum theory considered, this may mean
the state describes the entire universe at a given time
\footnote{We emphasize again that our discussion follows MGM in considering non-relativistic
  quantum mechanics, so this description has to be understood within a non-relativistic
  model.   In particular, the state does not involve gravitational
  degrees of freedom.}, with $\mathbb{C}^{d_n}$ (or some
$\mathbb{C}^{d_i}$ with $i>1$ in the case $n=\infty$) describing
an ``environment'' of effectively inaccessible degrees of freedom.
Alternatively,  $\mathbb{C}^{d_1} \otimes \ldots \otimes
\mathbb{C}^{d_n}$ may be a subsystem of the universe known
not to be entangled with the rest -- for example a collection
of systems prepared in a pure state, processed and distributed among laboratories
and kept isolated from the environment.
We consider measurements on system $1$ unless otherwise specified.

\subsection{Subjective and objective factorisations}

The alternative measurement postulates we consider below do not require the factorisation
$\mathbb{C}^{d_1} \otimes \ldots \otimes \mathbb{C}^{d_n}$ to be objective.  
Because there is no finite speed signalling bound in non-relativistic
quantum mechanics, in principle measurements can be carried out
arbitrarily swiftly on any degrees of freedom, whether
or not they are localized.
In the non-relativistic context it thus makes sense to consider
postulates (quantum or alternative) for
measurements that can similarly be applied to any factor of
any factorisation, whether or not it describes localized degrees
of freedom. There will generally
be many isomorphisms of the form 
\begin{equation}
\mathbb{C}^{d_1} \otimes \ldots \otimes
\mathbb{C}^{d_n} \simeq \mathbb{C}^{d'_1} \otimes \ldots \otimes
\mathbb{C}^{d'_p} \,  ,
\end{equation}
where $ \prod_{i=1}^n d_i = \prod_{i=1}^p d'_i $
and $\{ d_i \} \neq \{ d'_i \}$.   

However, it is worth noting that there {\it are} nonetheless
naturally preferred factorisations in non-relativistic
quantum mechanics, since laws defined in terms of one or more natural factorisations are
natural candidates for possible extensions of or alternatives to
quantum mechanics. Examples of natural factorisations include
\begin{equation}
  {\mathcal H}_{\rm total} = \bigotimes_{i=1}^k {\mathcal H}_{ i} \, ,
\end{equation}
where the index $i$ enumerates different particle types, or
particles of different masses $m_i >0$,
and 
\begin{equation}
  {\mathcal H}_{\rm total} =  {\mathcal H}_{\rm boson} \otimes
  {\mathcal H}_{\rm fermion} \, .
\end{equation}
It is interesting to consider the possibility that post-quantum measurement postulates 
we discuss below could apply (only) to these or
other specific factorisations or sets of factorisations.
Although we will not pursue
this further here, it also offers another natural way of defining
alternative measurement postulates, in which measurements can only
be applied to specific factors (for example bosonic or fermionic
degrees
of freedom) and the types of allowed
measurement depend on the factor to which the measurements apply.

\subsubsection{\bf Local Factorisations}

Another example is the factorisation defined by
degrees of freedom associated with different local regions.
In the idealized limiting case in which systems are
effectively pointlike, this gives us
\begin{equation}
  {\mathcal H}_{\rm total} = \bigotimes_{x \in \mathbb{R}^3}
  {\mathcal H}_{ x} \, .
\end{equation}

Coarse-graining gives
\begin{equation}
  {\mathcal H}_{\rm total} = \bigotimes_{i}
  {\mathcal H}_{V_i} \, ,
  \end{equation}
where the regions $V_i$ define a partition of $ \mathbb{R}^3$
into local regions, which could for example be defined by a regular lattice.

Alternative measurement postulates based on these factorisations
are particularly natural options when we consider extensions of non-relativistic quantum
mechanics that could fit well with special and general relativity.  
In this context, to respect the causal structure, it is natural
to take ${\mathcal H}_{\rm total}$ and its decompositions to define the state
on (or more precisely, asymptotically close to) the past light
cone.\cite{kent2005nonlinearity}
We do not discuss this further here, given MGM's focus on
non-relativistic quantum mechanics, but note
that it is one of the main motivations for
considering the possibility of alternative postulates
of the type we discuss in extensions of relativistic quantum theory \cite{kent2005nonlinearity,kent2021testing,kent2021quantum}. 

\section{Post-quantum measurement postulates} 

All the measurement postulates considered in this section are defined
via
hypothetical devices that give information
about a pure quantum state $\psi \in \mathbb{C}^{d_1} \otimes \ldots \otimes
\mathbb{C}^{d_n}$ of a set of $n$ systems.
To simplify the discussion 
we assume here the quantum state $\psi$ is not
altered by any of our postulated post-quantum measurements.
That is, we take the post-measurement update rule to be trivial,
although non-trivial alternatives (for example
those that define consistent non-linear versions of quantum theory \cite{kent2005nonlinearity}) are also interesting.  

The postulates apply whether $\psi$ is an entangled state or a product
state with respect to the given factorisation.
If $\psi$ is randomly drawn from an ensemble $ ( \psi_i , p_i )$,
the postulates apply to the state $\psi_i$ actually
chosen, whether or not this choice is known to the device user.
This last rule is consistent with the quantum measurement
postulate and seems the most natural option for an alternative
measurement postulate, since we do not expect measurement
probabilities or outcomes to depend on a user's knowledge.
It is consistent with MGM's treatment, as in Eqn. (4) of
Ref. \cite{masanes2019measurement}.   
We discuss the role of proper mixed states in MGM's argument further below.

\subsection{State readout devices}

An infinite precision {\it state readout device} $RD$ applied to
$\psi \in \mathbb{C}^{d_1} \otimes \ldots \otimes \mathbb{C}^{d_n}$
outputs an infinite precision classical description
of
\begin{equation}
\rho_1 = \Tr_{d_2 \ldots d_n} ( \ket{\psi} \bra{ \psi} ) \in O (
\mathbb{C}^{d_1} ) \, .
\end{equation}
Here $O (
\mathbb{C}^{d_1} )$ denotes the linear operators on
$ \mathbb{C}^{d_1} $. 
The operator $\rho_1$ defines the reduced density matrix of $\psi$ for
system $1$
and is normalised, hermitian and positive semi-definite.
The description is given as coordinates in some chosen basis, which we assume
is either input into the device prior to the readout or defined by
the construction of the device.   This description is output in
some idealized classical form, for example as an infinite printout or through a set of
pointer readings.  A finite set of pointer readings could suffice, given idealized
pointers whose positions are in principle readable with infinite
precision.   We discuss these idealizations further below.  

A finite precision state readout device $FPRD$ takes as input
a positive integer $m$.    Applied to
$\psi \in \mathbb{C}^{d_1} \otimes \ldots \otimes
\mathbb{C}^{d_n}$ it outputs an classical description
of the reduced density matrix (\ref{reddmone}) 
with respect to the given basis, 
to $m$ digit binary precision, in the sense that the coefficients
of basis elements are given to the nearest multiple of $2^{-m}$. 

\subsection{State function readout devices}

Let $f: O ( {\mathbb C}^{d_1} ) \rightarrow O ( {\mathbb C}^{d_1} )$
be a function mapping density matrices to density matrices.

An infinite precision {\it state function readout device} $FRD$ applied to
$\psi \in \mathbb{C}^{d_1} \otimes \ldots \otimes
\mathbb{C}^{d_n}$ outputs an infinite precision classical description
of $f ( \rho_1 )$ . 
The description is given as above.

A finite precision state function readout device $FFRD$ takes as input
some positive integer $m$ and outputs the
function value, in a prescribed basis, to $m$ digit binary precision.

\subsection{Expectation value readout devices}

An infinite precision {\it expectation value readout device} $ERD$
takes as input some hermitian
observable $A$ defined on system $1$, and outputs, to
infinite precision, the expectation value $\Tr ( A \rho_1 )$. 

A finite precision expectation value readout device $FERD$ takes as input
some positive integer $m$ and outputs  $\Tr (
A \rho_1 )$ to $m$ digit binary precision.

\subsection{Stochastic eigenvalue readout devices}

A {\it stochastic eigenvalue readout device}  $SEVRD$ takes as input
a hermitian observable $A$, with finitely or countably many
eigenvalues, defined on system $1$.
It produces as output
data that identifies an eigenvalue $\lambda_i$ of $A$, randomly chosen using the Born
probabilities $\Tr (P_i  \rho_1)$, where $P_i$ are the
projections onto the corresponding eigenspaces.
The output may identify the relevant eigenvalue, without
necessarily directly representing its value.   
For example, if the eigenvalues are labelled so that $\lambda_1 <
\lambda_2 < \ldots $, the possible outputs may be positive integers,
with the eigenvalue $\lambda_i$ identified by output $i$.

Note that these input and output rules are the same as those
for a quantum measurement of $A$.   However, an $SEVRD$ does
not alter the input quantum state. (Recall that we are
postulating a trivial post-measurement state-update.)
It can thus be applied
repeatedly to estimate the quantum probability distribution
for outcomes of measurements of $A$.   

A finite precision stochastic eigenvalue readout device $FSEVRD$ takes as
additional input some positive integer $m$ and outputs the relevant
eigenvalue to $m$ digit binary precision. 
The Born probabilities defining the random choice are still calculated to
infinite precision.

An integer labelled stochastic eigenvalue readout device $ISEVRD$ 
outputs the integer labelling the relevant eigenvalue, given some
ordering $\ldots \lambda_{-1} < \lambda_0 < \lambda_1 < \ldots$, or
$\lambda_0 < \lambda_1 < \ldots$, or   $\ldots \lambda_{-1} <
\lambda_0$ of the eigenvalues (if there are infinitely many descending
and ascending, not infinitely many descending and not infinitely many
ascending, respectively). 

A finite integer labelled stochastic eigenvalue readout device $FISEVRD$ 
takes additional input some positive integer $m$.  It uses some given 
ordering $\ldots \lambda_{-1} < \lambda_0 < \lambda_1 < \ldots$,
which may terminate in either direction or both.   
For $|i| \leq m$ it outputs the label $i$ with probability
$\Tr (P_i  \rho_1)$, where $P_i$ is the 
projection onto the corresponding eigenspace.
With probability $ ( 1 - \sum_{i=-m}^m \Tr (P_i  \rho_1) )$
it outputs an error code, for example, ``overflow''. 

\subsection{Stochastic state projection eigenvalue readout devices}

A {\it stochastic state projection eigenvalue readout device} $SPRD$ is a special case of a stochastic
eigenvalue readout device that takes an input a classical description of $P_{\phi}$, 
a projection onto a specified pure state $\phi \in
\mathbb{C}^{d_1}$.  It outputs $1$ with probability
$\Tr (P_\phi  \rho_1)$ and $0$ with probability
$\Tr ((1 - P_\phi )  \rho_1)$.

\subsection{Stochastic uncertainty readout devices}

A {\it stochastic uncertainty readout device} $SURD$ takes as input
some
hermitian observable $A$, with finitely or countably many eigenvalues,
defined on system $1$. 
It produces as output
data that identifies an eigenvalue $\lambda_i$ of $A - \langle A \rangle$, randomly chosen using the Born
probabilities $\Tr (P_i  \rho_1)$, where $P_i$ are the
projections onto the corresponding eigenspaces.

The output is the numerical value of the relevant eigenvalue, to
infinite precision.    
An $SURD$ could be constructed by applying an $ERD$ and 
$SEVRD$ in either order and combining their outputs, given
that neither device alters the quantum state.
We consider it, though, as a standalone device.

A finite precision stochastic uncertainty readout device $FSURD$
takes as additional input a positive integer $m$ and
outputs the relevant eigenvalue to binary precision $m$.
The Born probabilities defining the random choice are still calculated to
infinite precision.

\subsection{Stochastic positive operator devices}
A {\it stochastic positive operator readout device}  $SPOD$ takes as input
a complete finite or countable ($n = \infty$) set of positive operators $\{ A_i \}_{i=1}^n$, with $\sum_i A_i = I$.
It produces as output
data that identifies one of the $A_i$, randomly chosen using the Born
probabilities $\Tr (A_i  \rho_1)$.

Note that these input and output rules are the same as those
for a quantum measurement of the POVM $\{ A_i \}$.   However, an $SPOD$ does
not alter the input quantum state. (Recall again that we are
postulating a trivial post-measurement state-update.)
It can thus be applied
repeatedly to estimate the quantum probability distribution
for outcomes of measurements of $\{ A_i \}$.   

A finite integer labelled stochastic positive operator readout device $FSPOD$ 
takes additional input some positive integer $m$.
For $i \leq m$ it outputs the label $i$ with probability
$\Tr (A_i  \rho_1)$.
With probability $ ( 1 - \sum_{i=1}^m \Tr (A_i  \rho_1) )$
it outputs an error code, such as ``overflow''.

\subsection{State overlap devices}

A {\it state overlap device} $SOD$ takes a classical description of a
specified pure state $\phi \in
\mathbb{C}^{d_1}$ and real parameter $a \in (0,1)$ as inputs.
It outputs $1$ if  $\Tr (P_\phi  \rho_1) > a$ and $0$
otherwise.   

\subsection{Smoothed state overlap devices}

Various smoothed versions of state overlap devices can be defined.
As an illustrative example, we define a {\it smoothed state overlap
  device} $SSOD$ to take a classical
description of a specified pure state $\phi \in
\mathbb{C}^{d_1}$, a real parameter $a \in (0,1)$
and a real parameter $k>0$ as inputs.
It outputs $1$ with probability
$ 1 / (1 + \exp ( -k(  \Tr (P_\phi
\rho_1)   -a)))$ and
$0$ with probability 
$  \exp ( -k(  \Tr (P_\phi
\rho_1)   -a)) / (1 + \exp ( -k(  \Tr (P_\phi
\rho_1)   -a)))$.  

In the limit $k \rightarrow \infty$, this reproduces the behaviour
of an $SOD$.   

\subsection{Basis selection devices}

A {\it basis selection device} $BSD$ takes a classical description
of an orthonormal basis $ \{ \phi_i \}_{i=1}^{d_1}   $ of 
$\mathbb{C}^{d_1}$ as input. 
It outputs the value of $i$ that maximizes
$\Tr (P_{\phi_i}  \rho_1) $,
choosing randomly among maxima if there is more than one.  

\subsection{Smoothed basis selection devices}

Various smoothed versions of a basis selection device can be defined.
As an illustrative example, we define
a {\it smoothed basis selection device} $SBSD $ to take a classical
description of a 
specified orthonormal basis $ \{ \phi_i \}_{i=1}^{d_1}   $ of 
$\mathbb{C}^{d_1}$ and a real parameter $k>0$ as input. 
It outputs a value of $i$ chosen randomly from a distribution with probabilities  $N \exp
( k \Tr (P_{\phi_i}  \rho_1) )$, where $N$ is a normalisation
constant. 

In the limit $k \rightarrow \infty$, this reproduces the behaviour
of a $BSD$.

\subsection{Entropy meters}

An infinite precision {\it von Neumann entropy meter} $VNEM$ outputs the von Neumann
entanglement entropy
$ S( \rho_1) = - \Tr ( \rho_1 \log \rho_1 )$. 
An infinite precision {\it Renyi entropy meter} $REM ( \alpha) $, for real $\alpha \geq
0$,
with $\alpha \neq 1$, outputs the Renyi
entanglement entropy
\begin{equation}
S_{\alpha} (\rho_1) = \frac{1}{ 1- \alpha } \Tr ( ( \rho_1
)^{\alpha} ) \, .
\end{equation}

Writing the von Neumann entropy $S (\rho_1 )$ as $S_1 ( \rho_1 )$,
these can be combined to define an infinite precision {\it universal
  entropy meter} $UEM$, which takes as input a real $\alpha \geq 0$
and outputs $S_{\alpha} (\rho_1 )$.

Finite precision versions of these meters take as input
a positive integer $m$ and output the relevant
quantities to $m$ digit binary precision.

\subsection{Entropy certifiers}

A universal entropy certifier $UEC$ takes as input a real $\alpha \geq
0$ and a real $E > 0$.   If $S_{\alpha} (\rho_1 ) > E$ it outputs $1$, otherwise
$0$.

For $d_1 < \infty$, we have $S_{\alpha} (\rho_1 ) \leq \log (d_1 )$, with
equality if and only if $\rho_1$ is the uniformly mixed density
matrix $( { \frac{1}{d_1} }) I_{d_1} $ \cite{van2002renyi}.   We thus take the allowed
input range to be $0 < E
< \log ( d_1 )$ when $d_1$ is finite.   

Various smoothed versions of entropy certifiers can be defined.
As an illustrative example, we define a smoothed $UEC$ as one that
takes as input real
parameters $\alpha \geq 0$, $ E> 0$ and $k>0$. 
It outputs $1$ with probability
$ 1 / (1 + \exp ( -k(  S_\alpha ( \rho_1 ) - E)))$
and
$0$
with probability
$ \exp ( -k(  S_\alpha ( \rho_1 ) - E)) / (1 + \exp ( -k(  S_\alpha (
\rho_1 ) - E)))$.

In the limit $k \rightarrow \infty$, this reproduces the behaviour
of a $UEC$.

\subsection{Entanglement analysers}

An entanglement analyser $EA$ takes as input an orthonormal
basis
$\{ \psi_i : 1 \leq i \leq d_1 \}$ of $\mathbb{C}^{d_1}$.   It produces as output a matrix
$( M_{ij} )_{i=1,j=1}^{d_1 , d_1}$, whose elements are defined by
\begin{equation}
  M_{ij} =  \langle \phi_i | \phi_j \rangle \, ,
\end{equation}
where
\begin{equation}
  \langle \psi_i | \psi \rangle = | \phi_i \rangle 
\end{equation}
where the left hand side is a partial inner product
and $| \phi_i \rangle \in \mathbb{C}^{d_2} \otimes \ldots \otimes
\mathbb{C}^{d_n}$.

A finite precision entanglement analyser $FPEA$ takes as
additional input a positive integer $m$, and outputs the
matrix to $m$ digit binary precision, in the sense that the
real and imaginary parts of the matrix elements are given to the nearest multiple of $2^{-m}$.

\section{Comments on infinite precision inputs, outputs and probabilities}

\subsection{Inputs}

Most of the above devices are defined to have classical inputs that
are specified with infinite precision.  The readout devices require
a specification of a reference basis.   Expectation and eigenvalue
readout devices require a hermitian operator $A$ to be input;
state projection and overlap devices require a classical description
of a state $\phi$; in both cases, these need to be specified with respect to a reference basis. 
Universal entropy meters and certifiers require a real number
$\alpha$ as input.  Smoothed devices also require a smoothing
parameter, $a$ or $E$.

Infinite precision inputs cannot be typed on a keyboard in finite
time.  An analogue classical input, like a freely rotatable dial, can only be as precise as 
the approximation in which its classical description holds, i.e., only
finitely precise.  
So infinite precision inputs are not physical.

However, the same point applies to the standard measurement postulate of quantum
theory.   As given above, it requires each measurement outcome
to be represented by a linear operator.   If we think of a
measurement as a device with inputs and outputs, the inputs
are a set of linear operators $\{ Q_i \}$.  Specifying the $Q_i$ in general
requires infinite precision numbers with respect to a given
reference basis.   

The input prescriptions for our post-quantum measurement postulates
thus should be understood as idealizations similar to those 
of the standard quantum measurement postulate. 
A real world quantum measurement deviates from the ideal
in various ways, including that it is a 
process with some non-zero duration rather than an
instantaneous act.  For quantum measurements, the relevant operators, outcomes and 
probabilities are determined by the physics of 
the measurement devices.
There may be a sense in which they are
defined to infinite precision in nature, but we cannot determine
them to infinite precision.
If any of our hypothetical post-quantum measurements
were realised in nature, via presently unknown physics,
we should presumably expect the real world version to
similarly deviate from the ideal version.   
However, as in the quantum case, allowing infinite precision inputs in
the measurement postulates is a mathematically convenient idealisation
that does not, per se, 
necessarily represent new or problematic physics.  

\subsection{Outputs}

Infinite precision state readout devices, expectation
value devices and entropy meters are defined to give
infinite precision outputs.   Clearly, these too
are unphysical if taken literally.  Printing out
real numbers to infinite precision would take
infinite time and storage.  An analogue classical
output can only be finitely precise.

Again, analogous issues arise in the idealized version of standard
non-relativistic quantum mechanics found in many textbooks.
For example, one finds the statement that measuring the position $ \hat{\underline{x}}$ of a
single particle in state $\psi$ produces the
outcome $\underline{x}$ with probability density
$ | \psi ( \underline{x} ) |^2$.
The postulated outcomes here are vectors in $\mathbb{R}^3$, specified
to infinite precision.

To be clear, the example of infinite precision position measurements goes beyond the postulates
framed by MGM, which restrict to Hilbert spaces of
countable or finite dimension and to measurements
with finitely or (perhaps) countably many outcomes.\footnote{MGM
  describe a measurement as represented by operators
  corresponding to outcomes $Q_1 , \ldots , Q_n$ that
  satisfy the normalisation condition $\sum_{i=1}^n Q_i = I$.
  They do not explictly say that $n = \infty$, defining
  a countable sum, is allowed.  However, the
  measurements may be on spaces ${\mathbb C}^d$, where
  $d= \infty$, representing a countably infinite-dimensional space,
  is explicitly allowed by MGM, so one might assume so.}
Infinite precision position measurements also imply an
unnormalisable post-measurement state with infinite momentum
uncertainty and infinite energy. 
Nonetheless, defining quantum measurements
with infinite precision outputs is a useful idealization.
Real world position measurements have finite precision, of
course, and are treated via POVMs that involve a stochastic
smoothed version of the infinite precision postulate above.
The same would presumably be true of our hypothetical
post-quantum measurements, if realised in nature.
So, we regard the finite precision versions as
roughly analogous to an approximate position measurement
defined (for example) by Gaussian POVMs, in the sense that
they are more realistic and less physically problematic, although still idealized.

\subsection{Probabilities} 

The probabilistic versions of our post-quantum measurement
postulates involve probabilities defined as mathematical
expressions, which are typically real numbers defined
to infinite precision.
This is also true of the probabilities defined by
MGM's version (and other versions) of the standard quantum measurement
postulate.

There is no consensus on whether there is a fundamentally
satisfactory interpretation of probabilities
Whatever view one takes on this, one might also ask whether additional problems
arise from the fact that the probabilities in (post-)quantum theory
are defined to infinite precision.
For example, on a frequentist view, probabilities
are expressed as the frequencies in an infinite set of trials.
If a given probability has any physical representation in the universe,
this must thus involve an infinite number of events, which
would occur in infinitely many separate space-time regions,
spread over infinite space, time or both.
This might not be possible in our universe (or even a hypothetical
multiverse).
Unlike the hypothesis of an infinite precision output from
an effectively instantaneous localised measurement, though, it
does allow the possibility that infinite
resources can be available to express an infinite precision quantity.

In any case, we see no difference between the
issues raised by infinite precision probabilities in
standard quantum measurement postulates and in our hypothetical
post-quantum measurement postulates.

\section{Properties of OPFs and our alternative postulates}

Recall that our alternative measurement postulates
are defined for a set of $n$ physical systems
that have no entanglement with
any other systems, so that their state is a 
ray in $\mathbb{C}^{d_1} \otimes \ldots \otimes
\mathbb{C}^{d_n}$.
MGM's definition of a complete set of OPFs as one
satisfying properties $(1$-$3)$ above relies on their
tacit assumption that OPFs and measurements are defined by
their action on pure states.
In verifying that our alternative postulates satisfy
their postulates  $(1$-$3)$, we thus need only consider measurements on a pure
state $\psi \in \mathbb{C}^d$ of a single system, for
properties $1$ and $2$, and measurements on a product of pure states,
$ \psi \otimes \phi \in \mathbb{C}^d \otimes \mathbb{C}^b$
for property $3$.

Applying Properties  $(1$-$3)$ to our alternative measurement
postulates raises some technical issues.
For example, consider the infinite precision state
readout device.
This satisfies Property $2$: if we apply a unitary $U$ before
using the device, we get a readout of $U \psi$.
To test Property $3$, we need to apply it to
$ \psi \otimes \phi \in \mathbb{C}^d \otimes \mathbb{C}^b$.
This gives a readout of $\psi \otimes \phi$ in some chosen
basis.   If $\phi$ is known to infinite precision, then of
course in principle this defines $\psi$ to infinite
precision.  However, we also need to suppose that we are able to carry out
infinite precision calculations (ideally, in finite time)
to obtain $\psi$.
To test Property $1$, we need to consider random mixtures.
For example, if we apply unitary $U_i$ with probability $p_i$,
then forget which $U_i$ was applied, the readout produces
$U_i \psi$ with probability $p_i$ as required.
To test property 3 non-trivially on a random mixture of readout device and quantum measurements
we require a way of equating at least some of the outcomes.
For example, consider a mixture of one quantum measurement and one
readout device measurement.   Take the outcomes of the
quantum measurement to be positive integers.
We can then suppose some post-processing device
applied to the random measurement output that maps
the classical descriptions of a 
countable subset $\{ \psi_i \}_{i=1}^{\infty}$ of the pure states
to positive integers, taking $\psi_i$ to $i$.  (Again, ideally, this
should take finite time.)
Forgetting whether we carried out the quantum measurement
or the readout device plus post-processing then gives
mixtures of the positive-integer outcomes.

Finite precision readout devices avoid the idealisation
required in assuming that infinite operations require finite time.
However, a given version of finite precision generally
only approximately commutes with the operations involved in
Properties $(1$-$3)$.   For example, if $\psi_f$ is the state
described by a finite precision readout of the quantum
state $\psi$, and $U$ is a unitary, then $ U ( \psi_f ) \simeq ( U \psi
)_f$ but in general equality is not precise.

Another issue arises when we consider the relationship of our devices
to Property $3$.
Consider, for example, stochastic eigenvalue readout devices.
Let $A$ be a non-degenerate hermitian observable on $\mathbb{C}^d \otimes
\mathbb{C}^b$.
Suppose an $SEVRD$ with input $A$ is applied to 
$ \psi \otimes \phi \in \mathbb{C}^d \otimes \mathbb{C}^b$.
This produces one of the $db$ eigenvalues $\lambda_i$ of $A$
as output.  We can, following the statement of Property $3$, view this as a measurement on $\mathbb{C}^d$
parametrized by $A$ and $\phi$.  However, the measurement does not generally
correspond to any $SEVRD$ producing eigenvalues of some observable
$A_{\phi}$ on $\mathbb{C}^d$.   The projective measurement on the
joint system defines a positive operator valued measurement on
the first subsystem.   Hence the action of an $SEVRD$ on the joint system corresponds to the
action of a stochastic positive operator device on $\mathbb{C}^d$. 
The moral of this example is that, to be consistent with MGM's
postulates, we need to postulate measurements defined by $SPOD$s
rather than restricting to $SEVRD$s.

This issue applies to many of our devices: $ERD$s, $SPRD$s, $SURD$s,
$(S)SOD$s, $(S)BSD$s, $VNEM$s, $REM$s, $UEC$s, $EA$s.
While these have well defined actions on composite systems,
they do not generally reduce to the action of simple devices
on an individual subsystem.   

The simplest and most natural resolution of these various issues
is to extend our alternative measurement postulates to include
the closure (as defined by Properties  $(1$-$3)$ of the measurements
defined by the relevant devices.
Thus, we extend the category of $FPRD$s, $EA$s, and so on, to include devices defined by any finite sequence
of mixtures (including mixtures with quantum measurements),
composition with unitaries, and system composition.

The relevant probability distributions and outcome labellings,
unitaries, and specified extended systems with specified
background pure states are specified as additional
inputs, in the order in which they are applied.

\section{Relation of our post-quantum measurement postulates to standard quantum
  mechanics}

\subsection{Logical consistency and compatibility with relativity}

All of our post-quantum measurement postulates define ways of
obtaining information about unknown quantum states, without
disturbing them, that are not allowed by standard quantum theory.

Recall our convention that our
post-quantum measurements are carried out on system $1$.  If
\begin{equation}
  \psi = \psi_1 \otimes \ldots \otimes \psi_n
\in \mathbb{C}^{d_1} \otimes \ldots \otimes
\mathbb{C}^{d_n}
\end{equation}
is a product state, an infinite precision state readout device
gives a classical description of $\psi_1$, without altering $\psi$.
This allows an observer of system $1$, who does not know the state
prior to using the readout device, to construct a second copy of $\psi_1$, violating the
quantum no-cloning theorem \cite{WZ82,DIEKS1982271}.

A classical description of $\psi_1$ to arbitrary precision can
also be obtained by repeated uses of an infinite precision expectation device,
or a stochastic eigenvalue readout device, or a state overlap
device, or a smoothed state overlap device, with suitable
choices of operators and states as inputs.
A classical description of $\psi_1$ to high precision can also be
obtined by using an FFRD with suitably high precision, or by
repeated uses of suitably high precision versions of the other devices
just listed. 
Devices with these powers violate extensions \cite{PhysRevLett.79.2153,PhysRevLett.81.2598,10.1063/1.532887,PhysRevA.58.1827,PhysRevA.63.052313,Cerf01022000,10.5555/2011656.2011662} of the no-cloning theorem that bound the
sum of the fidelities attainable by any quantum operations
that produce two imperfect copies of an unknown input state.

Recall again that if $\psi$ is an entangled state of the first system with the rest, it
has a Schmidt decomposition of the form
\begin{equation} 
\psi = \sum_{i=1}^k ( p_i )^{\frac{1}{2}} \phi_i \otimes \chi_i  \, ,
\end{equation}
where $k>1$, the $p_i >0$, $\sum_{i=1}^k p_i =1 $,
the states $\phi_i \in {\mathbb C}^{d_1}$ are
orthonormal, and the states
$\chi_i \in \mathbb{C}^{d_2} \otimes \ldots \otimes
\mathbb{C}^{d_n}$ are orthonormal.
A standard quantum projective measurement onto a
basis of $\mathbb{C}^{d_2} \otimes \ldots \otimes
\mathbb{C}^{d_n}$ that includes the $\chi_i$ then
results in the post-measurement state
\begin{equation}
\psi'_i = \phi_i \otimes \chi_i
\end{equation}
with probability $p_i$.
Before the measurement, the output of an infinite precision Renyi entropy meter
\begin{equation}
  S_{\alpha} (\rho_1 ) = \frac{1}{1- \alpha } \log \Tr ( \rho_1^\alpha ) =
\frac{1}{ 1 - \alpha } \log ( \sum_{i=1}^k (p_i )^\alpha )> 0 \, , 
\end{equation}
where $\alpha \neq 1$.  The pre-measurement output of an infinite precision
von Neumann entropy meter is 
\begin{equation}
  S (\rho_1 ) = - \Tr (\rho_1 \log (\rho_1 )) = - \sum_{i=1}^k p_i \log
p_i > 0 \, .
\end{equation}

After the measurement, the meters output
\begin{equation}
  S_{\alpha} (\rho'_1 ) = 0
\end{equation}
and
\begin{equation}
  S( \rho'_1 ) = 0 \, .
\end{equation}

An observer of
system $1$ with an entropy meter can thus tell whether or not the
measurement has taken place on the remaining systems, without
any direct communication from those systems.
This violates the quantum no-signaling principle, according
to which the probability
\begin{equation}
  P( a_1 | \psi ; M_1 , M ) =  P( a_1 | \psi ; M_1 ) \, , 
\end{equation}
where the left hand side is the probability 
of outcome $a_1$ from a measurement $M_1$ on subsystem $1$ of
a composite system in state $\psi$, conditioned on measurement
$M$ being carried out on the other subsystems
and the right hand side is the 
probability unconditioned on $M$.
$RD$s, $FRD$s, $ERD$s, $SEVRD$s, $SPRD$s, $SURD$s, $SPOD$s, $(S)SOD$s, $(S)BSD$s,
$UEC$s and $EA$s also violate the quantum no-signaling principle.  

Some folk intuitions suggest that violating no-cloning
or quantum no-signaling necessarily creates a logical inconsistency.
This would imply that the relevant postulates cannot
consistently be added to standard quantum theory.
This is not correct, as we now explain. 

One intuition is that violating no-cloning introduces
nonlinearities into quantum theory that are incompatible with
the tensor product structure given by the composite systems
postulate, leading to inconsistencies.  A state readout machine
does indeed allow an observer to evolve an
initially unknown state using unitaries that depend
on that state, so that the final state does not
depend linearly on the initial state.
However, any
evolution that can be implemented by observers equipped
with readout machines could also, in principle, be 
implemented by observers who know the initial state
of a set of systems and who learn the outcomes of
all measurements on each system \cite{kent2005nonlinearity}.
Hence, there is no inconsistency in this form of
nonlinearity \cite{kent2005nonlinearity}. 

A related intuition is that allowing cloning or some form of signaling-via-measurement
necessarily breaks the delicate relationship between quantum mechanics
and special relativity.   Even if this were true, it would not affect
the logic of our critique, since MGM
consider quantum and alternative measurement postulates for
non-relativistic quantum mechanics.  But it is not true.
It {\it is} true that assuming {\it both} that the effects of measurements
propagate instantaneously in some reference frame, {\it and} also
assuming that instantaneous cloning is possible, gives a 
form of signaling-via-measurement that is
indeed incompatible with special relativity \cite{gisin1990weinberg}.
As noted earlier, though, the hypothetical devices we describe can
be kept consistent with special relativity and with the
quantum measurement postulates, if we define the devices
to act on localized subsystems and define the
reduced density matrix of a localized subsystem
as the trace of the full quantum state defined
on (asymptotically close to) the past light cone \cite{kent2005nonlinearity}. 
So defined, the devices are sensitive to the
effects of quantum measurements within the
past light cone, but not outside, so that the
causal structure of special relativity is respected.

In summary, while our various post-quantum measurement postulates
certainly have novel features, they can consistently be combined with quantum measurement
postulates to form a post-quantum theory.

\subsection{Signalling without carriers?}

Our post-quantum measurement postulates use
information about systems $2$ to $n$ to define the
outcomes and/or outcome probabilities of measurements
on system $1$.   They thus imply that global information determines what
happens in local measurements.  This may
seem unnatural.   But this much is also true of the quantum
measurement postulates.  
Suppose that A and B are separated, share an entangled state $\psi$, and
have a common rest frame whose coordinates both use. 
If B carries out a measurement of his subsystem, quantum
theory tells us his outcome
probabilities depend on the reduced density matrix
$ \Tr_A ( \ket{ \psi } \bra{ \psi} )$, a quantity that
depends on the global state $\psi$.   In fact, these
outcome probabilities are precisely the same as
those for our stochastic eigenvalue and positive
operator devices.
Moreover, quantum theory tells us that the outcome probabilities of any subsequent (with respect to
the shared time coordinate) measurement
of A's subsystem depend on B's result, which is not the case
for the stochastic eigenvalue and positive operator devices.
In this sense quantum theory appears in greater tension (although
still not actual conflict \cite{PhysicsPhysiqueFizika.1.195,bell1966problem,gisin1990weinberg} ) with relativistic locality than
a theory in which these devices define the only form of measurement.

However, as noted above, when quantum measurements are combined with
our devices, they allow signalling between subsystems.  This is not
possible in standard quantum theory.   According to our alternative
measurement postulates, the fact, or choice, of
quantum measurements on one subsystem generally influences the outcomes
and/or outcome probabilities of device measurements on other separated
subsystems.   This need not violate the relativistic no-signalling
principle when the devices are considered in a relativistic context
\cite{kent2005nonlinearity}.
However, it violates the quantum no-signalling principle, and
also conflicts with the intuition that transmitting information
requires a physical carrier.

That intuition itself deserves some
scrutiny.   For example, the ways in which information is transmitted
in quantum field theory and general relativity have (at least) stretched
our understanding of ``physical information carrier''.
Still, it remains a common intuition that there is ``something'' -- a perturbation
of space-time, or a dynamical quantum field -- mediating
interactions in these theories.

One possible response to this is to go beyond MGM's postulates and add a no-signalling
postulate to the basic principles of quantum mechanics.
Another is to accept that the intuition that information
requires a carrier may simply be wrong and
that our alternative measurement postulates are reasonable
as they stand.
A third option is to recognize that quantum mechanics is
only an effective non-relativistic theory, and even relativistic
quantum field theory is incomplete, so versions of
these theories with different or additional measurement
postulates need not necessarily be taken as theories in their final form.
If new measurement postulates do apply in nature, they
may ultimately be understood as part of a fundamental
theory with new mechanisms for carrying information.
For example, if our postulates play a role in some nonlinear
theory combining quantum theory and gravity, it might be
that the gravitational degrees of freedom carry more
information than we currently assume.   

If the aim is simply to understand the logical relationship
between various postulates of quantum mechanics, the first
option is interesting to pursue.
If, though, the aim is to understand whether there remain
plausible alternatives to quantum mechancs, the second 
option seems more reasonable.  Given all the initially
counter-intuitive features of quantum mechanics and of 
fundamental physics, it seems hard to mount a case 
for the no-signalling-without-carriers intuition as dogma.  
The third point fortifies this stance.

\section{Refutations of MGM's purported derivations}

\subsection{Refutation of MGM's derivation of the measurement
  postulate}

Consider, for definiteness, a theory in which the only possible
measurements are those defined by a finite precision von Neumann entropy
meter, applied to the $d$-dimensional system $1$, and the
combinations of these measurements that define closure under
Properties $(1$-$3)$, as described above. 

Since $0 \leq S(\rho_1 ) \leq \log (d_1 )$, the number of
possible outputs when precision $m$ is input is bounded by $ \lceil 2^m \log (d_1 ) \rceil + 1 $.

The $FPVNEM$ thus defines a full measurement with at most $\lceil 2^m
\log (d_1 ) \rceil +1 $
output probability functions $f_0 , f_{2^{-m}} , \ldots$, where
$ f_a $ defines the probability of output $a$.
For any pure state $\psi \in \mathbb{C}^d$, $f_0 ( \psi ) = 1$
and $f_a (\psi ) = 0 $ for $a \neq 0$.
This also holds if we consider the $FPVNEM$ acting on 
a pure state $\psi \otimes \phi \in \mathbb{C}_d \otimes \mathbb{C}_b
\simeq \mathbb{C}_{db}$,
as required by Property 3.  

Properties $(1$-$3)$ are all trivially satisfied.
The ``possibility of state estimation'' is also
trivially satisfied, since $f_0$ has value $1$
on any ensemble $(\psi_r , p_r )$ of states $\psi_r \in
\mathbb{C}^d$.

On a single system the von Neumann entropy meter defines
a trivial measurement: its OPFs can be represented in the
form of MGM's Eqn. (15), with
\begin{equation}
  f_0 (\phi ) = \langle \phi | I | \phi \rangle {\rm~and~}
  f_a (\phi ) = \langle \phi | O | \phi \rangle {\rm~for~} a \neq 0 \,
  ,
\end{equation}
where $I$ and $O$ are the identity and zero operator.
We can write this as
\begin{equation}
  f_a (\phi ) = \langle \phi | F_a | \phi \rangle \,  ,
\end{equation}
where $F_a = I$ if $a=0$ and $F_a = O$ if $a \neq 0$.

But on a system comprising two or more subsystems the $FPVNEM$ defines
non-trivial measurements, and its
OPFs do not satisfy MGM's Eqn. (16): it is not true
that
\begin{equation}
( f_a \star g_b ) (\psi ) = \langle \psi | F_a \otimes G_b | \psi \rangle
\, ,
\end{equation}
for any entangled $\psi \in \mathbb{C}^a \otimes \mathbb{C}^b$. 
Our theory thus satisfies all MGM's assumptions, but its
measurements are not described by their measurement postulate,
which requires all measurements to satisfy their Eqns. (15) and (16).
This refutes their purported derivation of the measurement postulate.

\subsection{Refutation of MGM's derivation of the post-measurement
  state-update
  postulate}

Even in a theory where all OPFs satisfy MGM's
Eqns. (15) and (16), so that their measurement
theorem characterizes measurements within the theory, the quantum post-measurement state-update rule
need not hold.

Consider a theory in which only
measurements defined by stochastic positive operator readout devices
and the combinations of these measurements
that define closure under Properties $(1$-$3)$, as described above, are
possible. 

Stochastic positive operator readout device measurements satisfy MGM's Eqns. (15)
and (16) and are characterized by their measurement theorem.  A readout
of operator $A_i$ has OPF $f_{A_i}$ which satisfies
\begin{equation}
  f_{A_i} ( \phi ) = \langle \phi | A_i | \phi \rangle \, \qquad{\rm and} \qquad (
  f_{A_i} \star g_{B_j} ) ( \psi ) = \langle \psi | A_i \otimes B_j |
  \psi \rangle \,
\end{equation}
for any $\phi \in \mathbb{C}^d$ and $\psi \in \mathbb{C}^d \otimes
\mathbb{C}^b$.

Stochastic positive operator readout device OPFs also satisfy the ``possibility of state
estimation'' assumption, for the same reason that general
quantum measurements do.   The probability of outcome $A_i$
given a SPOD measurement on an ensemble $ ( \psi_r , p_r )$
is $\Tr (A_i \rho )$, where the ensemble density matrix
$\rho$ is given by Eqn. (\ref{dmdef}). 
The equivalence classes of ensembles in this theory
thus correspond to density matrices.   Knowing the
value of a finite set of positive operator OPFs
suffices to identify the density matrix and henee
the value of all other positive operator OPFs,
as in standard quantum theory. 

However  after an SPOD measurement of $\phi$ the post-measurement
state remains $\phi$.   The state update after outcome $A_i$ can be
represented by the completely-positive map $\langle \phi | A_i | \phi
\rangle   I $, but this map is $\phi$-dependent.   There is no
$\phi$-independent completely-positive map representing the
state-update, for stochastic positive operator readout device
measurements defined by general POVMs $\{ A_i \}$, as required
by MGM's Eqns. (2) and (3).
This refutes MGM's purported derivation of the post-measurement
state-update postulate.

\section{Further problems with MGMs assumptions and arguments}

One might perhaps wonder whether the refutations above rely on
technicalities in MGM's assumptions that might reasonably be 
altered so as to exclude the above counterexamples and others.

For example, the first counterexample highlights that
MGM try to derive the quantum measurement postulates
from assumptions that apply only to
measurements on pure states and proper mixtures.
Their assumptions could be extended to apply
to measurements on improper mixtures.

The second counterexample highlights that MGM
do not assume that a sequence of measurements
can be considered as a single measurement.
The joint probabilities for a sequence of SPOD measurements
on an ensemble $( \psi_i , p_i )$ depend in general
on the individual states and probabilities, not
only on the density matrix (\ref{dmdef}).
So (\ref{dmdef}) does not represent
an equivalence class of ensembles for
sequential measurements.   If these
were considered to be single measurements,
the argument above for the  
``possibility of state estimation'' 
would no longer hold.

However, there are several problems with this sort of "rescue
strategy".   

First and foremost, from a mathematical perspective, entropy meters, SPODs, and our other
hypothetical devices give well-defined and natural
measurement rules.   A classical physicist who knew nothing about quantum
theory except the structural postulates defining the space
of states, and was told that information could be obtained
about the states without altering them, would not be
surprised: essentially the same is true in classical physics.\footnote{
They might perhaps suspect that, as in classical physics, the statement
describes idealized measurements, and that a real world measurement
will always have a nonzero effect.   But they might reasonably
also assume that, as in classical physics, the effect can be made
arbitrarily small, and so the idealization is justifiable.}

They would also not be surprised to be told that ideal measurements
can be described by pure state readout devices, since again
classical physics suggests that states are ontic and directly
observable.   
Several of our other devices, including SPODs, can be built
from pure state readout devices with post-processing, so 
these too would not seem surprising.

Nor should they necessarily be surprised to be told that entropy
meters represent a possible type of measurement in quantum theory.
Having understood the structure of the quantum state space, they
would recognise that entanglement is a well-defined and quantifiable
feature of quantum states, with some nice properties, including
that entanglement measures are invariant under local unitaries.
It does not seem a huge leap to imagine 
it might plausibly represent a measurable physical quantity.

Second, it is not obvious that there {\it is} a sensible way to
extend the postulates to exclude both our counterexamples.
For instance, the obvious extension to address the first counterexample
is to reframe the assumptions to apply to 
measurements on a subsystem (say subsystem $1$) of
pure states in the space $\mathbb{C}^{d_1} \otimes \ldots \otimes
\mathbb{C}^{d_n}$ that represents a composite of $n$ systems.
But the finite precision von Neumann entropy meter
still satisfies Properties  $(1$-$3)$ in this setting.
It has a finite set of OPFs.   So, even if we consider
measurements on ensembles of the form $( \psi_r , p_r )$,
where $\psi_r \in \mathbb{C}^{d_1} \otimes \ldots \otimes
\mathbb{C}^{d_n}$, it trivially also satisfies the
``possibility of state estimation'' assumption,
since knowing the value of its OPFs on an ensemble determines the value
of any OPF (as there are no other OPFs in this counterexample).

Third, there is little or no intellectual value in mechanically 
extending a system of postulates until one excludes all known alternative
measurement rules.   We would not gain much insight from
showing that the quantum measurement postulates are
derivable from a set of rules specifically tailored to exclude
any others.  We need to look at least as critically at
any proposed set of assumptions as at any proposed alternative
measurement rules.   And in fact, even MGM's existing assumptions
make problematically arbitrary choices
that are hard to motivate.   We review these next.  

\subsection{Measurements with infinitely many outcomes}

MGM's Eqn.~(\ref{fullmment}) does not clearly distinguish finite and
infinite sums.    As written, it appears to restrict
full measurements to those with finitely many outcomes
(as does the definition of full measurement for quantum
mechanics, given after Eqn. (1) of Ref. \cite{masanes2019measurement}).
On the other hand, MGM allow the dimension $d$ of
a system's state space to take the value $d=\infty$,
representing countable infinity.   It would thus be
consistent to allow $n=\infty$, again representing
countable infinity, in Eqn. (\ref{fullmment}),
since standard textbook quantum mechanics certainly allows this.

Even this, though, is restrictive.  
For example, our state readout devices, expectation value
readout devices and entropy meters all have uncountably
many possible outcomes, even for states in finite-dimensional
spaces.  Standard quantum theory also allows POVMs with
uncountably many outcomes in finite-dimensional spaces. 

It might be argued that this feature of the relevant devices
is an artefact of the arguably
unphysical assumption that measurement outputs can be
given with infinite precision.
Similarly, it might be argued that position measurements and
other quantum measurements with uncountably many outcomes
are unphysical idealisations.   
So, it might be argued that it is reasonable for a formalism for measurement
postulates to restrict to countably many outcomes, as MGM appear to.
There is even a case for restricting to finitely many outcomes,
on the grounds that no realistic measurement can produce more
than finitely many different results in a given finite time interval.

In summary, MGM are unclear whether they intend to allow
infinite outccome measurements. Their assumptions could
be amended so as to explicitly exclude our infinite
precision devices.   We have
defined finite precision versions of all these devices because there is an arguable physical
case that any quantum or alternative measurement postulate
should describe finite precision measurements.  

\subsection{Definition of measurements via action on pure states}

MGM's definition of an OPF has a much more serious problem. 
It tacitly assumes that an OPF, and
hence a full measurement, is determined by its
action on pure states.
As the post-quantum measurement
postulates above illustrate, this is neither necessary
nor especially natural.   For example, the state
function readout devices corresponding to the
functions $f ( \rho_1 ) = \rho_1^n$, for positive integer $n$,
all produce the same output
\begin{equation}
  (\rho_1 )^n = \rho_1 = \ket{\psi_1} \bra{\psi_1}
\end{equation}
for
an input pure state $| \psi_1 \rangle$, but produce different
outputs when $\rho_1$ is an improper mixed state.
Similarly, universal entropy meters all produce output $0$ when
a pure state is input, but different outputs (depending
on the input $\alpha$) when $\rho_1$ is an improper mixed state.
Our other examples either have similar properties or can be
generalized so as to.

MGM do not justify their assumption that any conceivable type of 
measurement is necessarily determined by its action
on pure states.   We find it hard to see any good justification.
Entanglement is a fundamental
feature of the quantum formalism, whose existence follows directly
from the structural postulates above.  A purportedly
theory-independent characterization of measurements
ought to allow for measurements that are sensitive
to entanglement in ways that standard quantum
measurements are not.   One might perhaps hope to
rule this possibility out from other plausible
assumptions, but ruling it out by fiat is surely unsatisfactory.

\subsection{Problems with MGM's ``possibility of state estimation'' assumption}

\subsubsection{\bf The concept of ``state estimation'' depends on
the measurement postulates}

The logic behind MGM's characterization of their assumption appears
to be as follows.  Suppose we have a list of outcomes $f^1 , \ldots ,
f^k$ with the property MGM claim is required for state estimation,
i.e., that knowing their value on any ensemble
$( \psi_r , p_r )$
allows
us to determine the value of any other OPF $g \in
{\mathbb F}_d $ 
on the
ensemble $( \psi_r , p_r )$.
We can then generate a finite list of full measurements
$M^1 , \ldots , M^k$, where $M^i$ has outcomes $f^i , g^i_1 , \ldots ,
g^i_{k_i}$ and
\begin{equation}
  f^i + \sum_{j=1}^{k_i} g^i_j = I \, .
\end{equation}
(It is possible that $g^i_j = f^k$ for some values of $i,j,k$;
it is also possible that $M^i = M^j$ for some values of $i,j$.
This does not affect our argument, though it may make some of 
the measurements in the procedure
described below redundant.)    

Carrying out measurement $M^i$ repeatedly (say $N^i$ times) on a given
ensemble $( \psi_r , p_r )$
allows a statistical estimation 
of the probability $P_i = \sum_r p_r f^i ( \psi_r )$ of
obtaining outcome $f^i$ on the ensemble.
Knowing these probabilities precisely for all the $f^i$ would allow 
a precise determination of the value of any other OPF $g$ on the
ensemble, and hence would determine the equivalence class of the
ensemble.
Hence one might think that knowing the probabilities to good precision allows
the value of any other OPF $g$ on the ensemble to be
estimated to good precision, and hence determines the
equivalence class of the ensemble to good precision.

But does this last point follow?   In standard quantum theory,
we can indeed sketch an argument for it, as follows.

Knowing the $P_i$ precisely
determines the value of any OPF on the ensemble
$( \psi_r , p_r )$, and hence determines the
density matrix $ \rho = \sum_r p_r \ket{\psi_r } \bra{ \psi_r }$
that determines the equivalence class of the ensemble.
We know that density matrices in $d$ dimensions can
can be described with $k= (d^2 -1 )$ parameters.
Moreover we know that $P_i = \Tr ( \rho f^i )$ depends
linearly on $f^i$.   
Without loss of generality (removing redundant outcomes
$f^j$ from the given list if necessary if the list is
linearly dependent), we can thus
take $k= (d^2 -1 )$.

Standard statistical tests should, with high confidence,
give an estimate in the form $ P^e_i - \epsilon_i < P_i < P^e_i +
\epsilon_i$, where $P^e_i$ is the frequency of the outcome
corresponding to $f^i$ in the $N^i$ trials, and $\epsilon_i$ is
a suitable multiple of the standard deviation.
The vector function
\begin{equation}
  \underline{P} ( \rho ) = ( \Tr ( \rho f^1 ) \,
\ldots \,  \Tr ( \rho f^k ))
\end{equation}
           is a continuous,
differentiable, linear and invertible function of $\rho$.
Hence, assuming all the statistical estimates are valid (which will be
true with high confidence given suitable choices of the
$\epsilon_i$) restricts $\rho$ to a subset $B$ of the
set of density matrices on ${\mathbb C}^d$,
where $ F(\rho ,\sigma )= \Tr ( {\sqrt
      {{\sqrt {\rho }}\sigma {\sqrt {\rho }}}} )^{2} > 1 -
\epsilon$ for all  $ \sigma \in B $ and the parameter $\epsilon$
can be made arbitrarily small with suitably large $N^i$.
Thus we can obtain an estimate $\sigma$ of $\rho$ with fidelity $F(\rho, \sigma )$ arbitrarily
close to $1$, and with arbitrarily high confidence, by choosing the
$N^i$ to be sufficiently large.   

Several of these points need not necessarily hold true for
alternative measurement postulates, however.
Suppose some alternative measurement postulates hold, and we
are given a list of measurement outcomes $f^1 , \ldots , f^k$ with
the specified property: 
i.e. that  knowing their value on any ensemble
$( \psi_r , p_r )$
allows
us to determine the value of any other OPF $g \in
{\mathbb F}_d $ 
on the
ensemble $( \psi_r , p_r )$.

We can again generate a finite list of full measurements
$M^1 , \ldots , M^k$, where $M^i$ has outcomes $f^i , g^i_1 , \ldots ,
g^i_{k_i}$ and
\begin{equation}
  f^i ( \psi ) + \sum_{j=1}^{k_i} g^i_j (\psi ) = 1
\end{equation}
for all pure states $\psi$.
For an ensemble  $( \psi_r , p_r )$, this gives
us
\begin{equation}
\sum_r p_r ( f^i ( \psi_r ) + \sum_{j=1}^{k_i} g^i_j (\psi_r )) = 1 \, . 
\end{equation}

We write
\begin{equation}
  P_i = \sum_r p_r f^i (\psi_r ) \, .
\end{equation}

By assumption, knowing the $P_i$ precisely
determines the value of any OPF on the ensemble
$( \psi_r , p_r )$.
Carrying out measurement $M^i$ repeatedly (say $N^i$ times) on a given
ensemble $( \psi_r , p_r )$
allows us to make an estimate $P_i^e$ 
of the probability $P_i = \sum_r p_r f^i ( \psi_r )$ of
obtaining outcome $f^i$ on the ensemble.
Since knowing the probabilities $P_i$ precisely would determine the value of any other OPF $g$ on the
ensemble, it would determine the equivalence class (call it $E$) of the
ensemble.

However, it does not necessarily follow that knowing the probabilities to good precision allows
the value of any other OPF $g$ on the ensemble to be
estimated to good precision, and hence determines the
equivalence class of the ensemble to good precision.
We cannot (within MGM's framework) assume that the equivalence class of the ensemble
is represented by the
density matrix $ \rho = \sum_r p_r \ket{\psi_r } \bra{ \psi_r }$,
nor that it can be described by $k= (d^2 -1 )$ or any other
specific number of parameters, nor even that it is a closed or
connected subspace in ${\mathbb R}^n$ for some $n$.
We cannot assume that
\begin{equation}
  \underline{P} ( (\psi_r , p_r ) ) = ( f^1 ( ( \psi_r , p_r ) ) \,
\ldots \,  f^k ( ( \psi_r , p_r ) ) )
\end{equation}
is a linear, differentiable or
continuous function on the space of equivalence classes,
nor even that the space is such that these terms are well defined.   
We cannot assume that the $P^e_i$ correspond to any valid
equivalence class of ensembles.  Even if they do correspond
to an equivalence class $E'$, we
cannot assume that there is some fidelity surrogate
$FS$ defined on the equivalence classes, with the
property that $FS(E', E) \rightarrow 1$ as the
$N^i \rightarrow \infty$.

In short, under alternative measurement postulates, there need not
necessarily be any notion of state estimation resembling that
implied by quantum measurement postulates.
Even if there is such a notion, MGM's ``possibility of state
estimation'' assumption does not necessarily
imply that state estimation is possible.

\subsubsection{\bf State estimation is possible even when MGM's
  ``possibility of state estimation''  assumption fails}

\paragraph{State estimation with infinite precision readout devices}
\hfill\break
\hfill\break
Consider a system described by the finite-dimensional space
$\mathbb{C}^d$.   Suppose that the quantum measurement
postulate holds, as does the post-quantum
measurement postulate defined by some version of the state readout devices
discussed above.   Suppose that no types of measurement other
than quantum measurements and state readout measurements are possible.

Suppose first that we allow infinite precision state readout
measurements.   These define an uncountably infinite set
of OPFs $\{ f_{\psi} : \psi \in  \mathbb{C}^d \}$, defined
on pure states $\phi \in  \mathbb{C}^d $ by
\begin{equation} \label{purereadoutopf}
  f_{\psi} ( \phi ) =
\begin{cases}
    1 & \text{if  $\phi= \psi$} \, , \\
    0 & \text{otherwise} \, . 
\end{cases}
\end{equation}

For an ensemble $E = (\psi_r , p_r )$ this gives
\begin{equation}
  f_{\psi} ( E ) = \begin{cases} p_r & \text{if $ \psi = \psi_r$
      ~for~some~}$r$   \, , \\
    0 &  \text{otherwise} \, .
\end{cases}
\end{equation}

We also know there is a finite list of quantum measurement outcomes $f^1 , \ldots , f^{d^2
  - 1 }$ whose values on any ensemble $E = (\psi_r , p_r )$ determine
the value of all quantum measurement outcomes on that ensemble.

If the ensemble is a single pure state, $E= ( \phi , 1 )$, then the
values of $f^1 , \ldots , f^{d^2  - 1 }$ determine $\phi$.
They thus also determine the values of all the OPFs $f_{\psi}$,
from (\ref{purereadoutopf}).    If MGM's ``possibility of
state estimation'' assumption applied only to pure states,
rather than ensembles, it would thus be satisfied.

However, for a general ensemble $E = (\psi_r , p_r )$, the
values of $f^1 , \ldots , f^{d^2  - 1 }$ determine the
density matrix $\rho = \sum_r p_r \ket{ \psi_r } \bra{ \psi_r }$
but not the specific ensemble $E$.  Suppose that $\rho$ has
maximal rank $d$.
It is the density matrix of (infinitely) many different ensembles,
and every pure state $\psi$ belongs to some but not all of these
ensembles.\footnote{In fact, every $\psi$ does not belong to a
  generic finite or countable ensemble represented by $\rho$.}
Define $p_F (\psi ) = p$ if the ensemble
$F$ includes $(\psi, p )$ (i.e. includes the state $\psi$
with probability $p$), and $p_F ( \psi) = 0$ otherwise.
For every pure state $\psi$, the values $p_F ( \psi)$, for
ensembles $F$ whose density matrix is $\rho$, 
range over a finite interval $[0 , p_F^{\rm max} ( \psi ) ]$.
The values of $f^1 , \ldots , f^{d^2  - 1 }$ on $E$ thus do not
determine the values of any of the $f_{\psi}$ on $E$.

Now consider any finite list of outcomes that includes
the quantum measurement outcomes $f^1 , \ldots , f^{d^2  - 1 }$
together with some list $f_{\phi_1} , \ldots , f_{\phi_l}$ of
our post-quantum measurement outcomes.
A generic finite ensemble $E = (\psi_r , p_r )$ will
include none of the $\phi_i$, so that $f_{\phi_i} (E) = 0$
for all $i$.   There are infinitely many finite ensembles
$E'$ with the same density matrix $\rho$ as $E$ that
also include none of the $\phi_i$.   The values of
$f^1 , \ldots , f^{d^2  - 1 }$ and $f_{\phi_1} , \ldots , f_{\phi_l}$
do not distinguish among these ensembles.
Hence there is no finite list of outcomes whose
value on an ensemble $E$ allows us to determine the
values of generic $f_{\psi}$ on $E$.   
That is, MGM's ``possibility of state estimation'' assumption
fails for this combination of quantum and post-quantum
measurement postulates.

There are two significant issues here.   First, as noted
earlier, while the ``possibility of state estimation''
fails for ensembles, it holds for pure states.
The reason that it fails for ensembles is that,
because a state readout device allows ensembles
(not just density matrices) to be distinguished,
MGM's version of ``state estimation'' requires
the description of an ensemble to 
be inferrable from finitely many state readout outcomes,
which is impossible.
It is not clear, though, that the requirement is reasonable.
It is reasonable to base a measurement postulate on
the measurement of pure states, which play a fundamental role in any theory based
on the quantum formalism.
It is not obvious why a measurement postulate should be based on
measurements of proper mixtures, which need not necessarily have any
fundamental status.
For the quantum measurement postulate, the distinction does not
matter: if a finite set of measurement outcomes determines the
value of all quantum measurement outcomes on all pure states,
it also does on all mixed state density matrices.
MGM's approach either assumes the distinction will not matter
for any possible alternative measurement postulate (which we
have seen is false) or assumes a fundamental role for proper
mixtures without discussion or clear justification.

Second, even if we accept, for the sake of discussion, that
it is reasonable to base a measurement postulate on the
implications for measuring proper mixtures, it seems
wrong to suggest that the postulate characterizes the
possibility of state estimation.    Specifically, it
seems wrong to suggest that state estimation is not
possible in the example under discussion, which
combines quantum measurements with infinite precision
state readout devices.
We know that quantum measurements suffice to estimate
the density matrix $\rho$ of an proper mixture, in the
sense that, by carrying out sufficiently many
quantum measurements, we can obtain an estimate
$\rho_e$ such that (with very high probability)  $F( \rho_e , \rho ) > 1 - \epsilon$,
for any given $\epsilon > 0$.
But also, if we apply an infinite precision state
readout device repeatedly to a finite
ensemble $E= (\psi_i , p_i ) $, we can obtain an
estimate $E_e = (\psi^e_i, p^e_i )$, where
$ \{ \psi^e_i \} \subseteq \{ \psi_i \}$.
Write $p^e_i = 0$ if $\psi_i$ is not included
in the states of $E_e$.
If we repeat the operation sufficiently often,
we can ensure (with very high probability) that  $ \sum_i | p_i - p^e_i | < \epsilon$,
for any given $\epsilon > 0$, and hence that all outcome probabilities
can be estimated to within $\epsilon$.
This is a natural criterion for estimating a proper mixed state
defined
by a finite ensemble,
paralleling the procedure and result obtainable for quantum
measurements.

A similar discussion applies to estimating a countably infinite
ensemble $E=(\psi_i , p_i )_{i=1}^{\infty}$.   For any $\epsilon >0$,
we can ensure that (with arbitrarily high probability) all outcome probabilities $p_i > \epsilon$
are estimated to within (say) $\epsilon /2$, by repeating the readout device operation
sufficiently often.  This gives us an estimate for $E$ in the form of
a finite ensemble $E_e = (\psi^e_i,
p^e_i )$, where $p^e_i > 0$ and $ \{ \psi^e_i \} \subseteq \{ \psi_i
\}$, with the properties that (i) $\psi_i \in E_e$ if $p_i > \epsilon$
and (ii) $| p^e_i - p_i | < \epsilon/2$.   This is a natural criterion 
for estimation of a proper mixed state defined by a countably infinite ensemble. 
\hfill\break
\paragraph{State estimation with finite precision readout
  devices}
\hfill\break
\hfill\break
A similar analysis applies if we consider quantum measurements
together with (only) the post-quantum measurement postulate
defined by finite precision readout devices.

Suppose we apply a finite precision readout device repeatedly to
a finite ensemble $E=(\psi_i , p_i )_{i \in I}$, and set the precision
to increase suitably (for example, by one for each measurement).
We can then obtain an estimate $E_e = (\psi^e_i, p^e_i )_{i \in I'}$, where
$I' \subseteq I$.   Again we write $p^e_i = 0$ if $i \in I \setminus
I'$.   Then, if we repeat the operation sufficiently often,
we can ensure (with very high probability) that $ | \psi^e_i - \psi_i
| < \epsilon$ for all $i \in I'$ and that $ \sum_i | p_i - p^e_i | < \epsilon$,
for any given $\epsilon > 0$.
Hence the states in the ensemble with probability larger than
$\epsilon$, and their probabilities,
can both be estimated to within $\epsilon$.
This is a natural criterion for proper mixed state estimation
in a finite precision context, again paralleling the procedure
and result for quantum measurements.

As above, a similar discussion applies to estimating a countably
infinite ensemble.   
\hfill\break
\paragraph{State estimation with state overlap 
  devices}
\hfill\break
\hfill\break
For $\epsilon $ close to $1$, a state overlap device $SOD ( \phi ,
\epsilon )$ reports whether an input pure state $\psi$ is
close to $\phi$ in the sense that $\Tr ( P_\phi P_\psi ) = | \braket{  \phi}{ \psi} |^2 > 1 - \epsilon$.   For any given $\epsilon$, we can find a finite set
of $\psi_i$ such that for any pure state $\psi$ at least one $\psi_i$
satisfies $  | \braket{ \psi_i }{ \psi} |^2 > 1 - \epsilon$.
Hence, by repeated use of state overlap devices on a finite
ensemble $E= (\psi_i , p_i ) $, with suitable
choices of $\phi$ and with values of $\epsilon$ tending to $1$,
we can obtain an estimate $E_e = (\psi^e_i, p^e_i )_{i \in I'}$,
in a similar way to that obtained by finite precision readout
devices, and satisfying the same estimation criterion.

A similar analysis applies to smoothed state overlap devices.

\hfill\break
\subsubsection{\bf Devices satisfying the ``possibility of state
  estimation''}

\paragraph{State estimation with entropy meters}
\hfill\break
\hfill\break
Consider again a system described by the finite-dimensional space
$\mathbb{C}^d$.   Suppose now that the quantum measurement
postulate holds, as does the post-quantum
measurement postulate defined by one of the entropy
meters discussed above.  Suppose that no types of measurement other
than quantum measurements and entropy meter output measurements are
possible.

This example illustrates again the unnaturality of (i) assuming that
measurement outcome functions are uniquely defined by their action
on pure states, and (ii) considering their action on proper
mixed states in defining an assumption intended to characterize
the possibility of state estimation.
Given an ensemble $( \psi_r , p_r )$ of pure states in
${\mathbb C}^d$, an entropy meter outputs the entropy of
the state actually presented.  Since that state is one
of the pure states $\psi_r$, the meter always outputs $0$.
The ``possibility of state estimation'' assumption is thus
satisfied, since a list of $(d^2 - 1)$ quantum measurement
outcomes suffices to determine the output of all quantum
measurement outcomes on a given ensemble, and also (trivially)
determines the output of an entropy meter.  

Similar comments apply to entropy certifiers and smooth
entropy certifiers.  A universal entropy certifier applied
to an ensemble $( \psi_r , p_r )$ of pure states in
${\mathbb C}^d$ produces output $0$.
A smoothed
$UEC$ produces output $1$ with probability $
 1 / (1 + \exp ( k E)))$
and output $0$ with probability $ \exp (kE) / (1 + \exp ( k E)))$. 
\hfill\break
\paragraph{State estimation with stochastic eigenvalue readout
  devices}
\hfill\break
\hfill\break
Consider again a system described by the finite-dimensional space
$\mathbb{C}^d$.   Suppose now that the quantum measurement
postulate holds, as does the post-quantum
measurement postulate defined by the stochastic eigenvalue devices
discussed above.   Suppose that no types of measurement other
than quantum measurements and stochastic eigenvalue readout measurements are
possible.

The outcomes and outcome probabilities for a stochastic eigenvalue readout device
with input $A$ (a hermitian operator) are the same as those for a
quantum measurement of the observable $A$.   Hence they are determined
by any list of $(d^2 -1 )$ quantum measurement outcomes that suffices
to determine the proper mixed state $\rho_1$.
This example thus satisfies MGM's ``possibility of state estimation''
assumption.
It also satisfies a version of the assumption that applies to
improper mixed states.   
\hfill\break
\paragraph{State estimation with expectation value readout
  devices}
\hfill\break
\hfill\break
Consider again a system described by the finite-dimensional space
$\mathbb{C}^d$.   Suppose now that the quantum measurement
postulate holds, as does the post-quantum
measurement postulate defined by the expectation value devices
discussed above.   Suppose that no types of measurement other
than quantum measurements and expectation value output measurements are
possible.

For any observable $A$, the output of an expectation value device, $\Tr (A \rho_1 )$, is
determined by the proper mixed state $\rho_1$.
Hence it is determined
by any list of $(d^2 -1 )$ quantum measurement outcomes that suffices
to determine the proper mixed state $\rho_1$.
This example thus satisfies MGM's ``possibility of state estimation''
assumption.
It also satisfies a version of the assumption that applies to
improper mixed states.   

\section{Discussion}

We have highlighted several problems with MGM's analysis.
The most salient point is that their claimed derivation of
the quantum measurement postulates from structural postulates
is incorrect.   Their approach also has
other theoretically significant defects.
They ignore the crucial role of entanglement
in quantum theory by assuming that measurement postulates
can be framed in terms of the action of measurements on pure
states of a subsystem.  They assume a fundamental role for proper
mixed states, i.e., probabilistic ensembles of pure states,
in formulating quantum theory and defining measurement postulates.
They assume that, in any alternative to quantum theory,
proper mixed states must fall into large equivalence classes
(analogous to, although not necessarily taking the form of,
equivalence classes defined by mixed state density matrices
in quantum theory).   They assume that any alternative to
quantum theory must satisfy a particular property that
allows state estimation in quantum theory, even though
this property is not generally aligned with natural
notions of state estimation in general theories.  
Each of these assumptions restricts the allowed types
of measurement towards standard quantum measurements.
None of them is reasonable in an open-minded analysis
of the possibility of alternatives.

Of course, standard quantum measurement theory has some elegant
features.  Quantum measurements on a subsystem can generally be
treated, up to a point, as unitary quantum evolution of
a larger system.  This allows the Heisenberg cut to
be shifted at will, at least in a range between mesoscopic subsystems and conscious
observers, without significantly affecting the predictions.
This has even encouraged some to try eliminating measurement postulates
altogether via some Everettian approach
\cite{everett1957relative,dewitt2015many,saunders2010many}.
Critics have pointed out several problems with Everettian ideas (see
e.g. \cite{saunders2010many}
for some discussion).   
However, after nearly a hundred years, we have no empirical evidence for any alternative to standard
quantum measurements.

Still, standard quantum measurement theory has problems.
We seem to see definite measurement results, which suggests that the
Heisenberg cut cannot be shifted beyond the point of conscious
observation: hence the quantum measurement problem.
Another and perhaps sharper way of highlighting this issue is that
we frame quantum theory as a mathematically precise dynamical
theory, and understand it as a probabilistic theory, but have no
mathematically precise definition of its sample space.
We do not know if gravity is quantum.  If not, quantum theory
has a definitely limited domain.  A theory that somehow
combines a classical description of space-time with
quantum matter would necessarily imply, inter alia,
that space-time defines some forms of measurement
on quantum matter.  Such measurements need not
necessarily be standard quantum measurements.

There are also puzzles elsewhere, most notably in
cosmology, which suggest we may not fully understand
the quantum evolution of the universe, or possibly
that standard quantum theory does not describe its
evolution (e.g. \cite{kent1998beyond,kent2013beable,kent2022hodology}).

And then there is consciousness. 
Clearly consciousness involves some form of quantum
measurement: we consciously access information about quantum
systems.  Precisely how this works is mysterious, as is
the whole relationship between consciousness and physics
(see e.g. \cite{chalmers1996conscious,iepconscious,sep-consciousness}
for
some discussions).
There are also tensions between some arguably
natural-seeming assumptions about conscious perceptions
and quantum theory (see e.g. \cite{wigner1961remarks,frauchiger2018quantum,brukner2018no,bong2020strong,
Wiseman2023thoughtfullocal}).  

None of these puzzles show that the standard quantum
measurement postulates are definitely inadequate.  Nor do they
point to specific alternative postulates that would
definitely resolve them, as far as we can currently see.
But they do make a clear case for keeping an open mind,
and for spending more time looking for interesting
empirically testable alternatives to quantum theory and less on efforts
to show it is essentially unique.

\section{Responses and Comments}

As the abstract notes, this paper responded to 
Masanes-Galley-M\"uller's original article \cite{masanes2019measurement}.
It produced a response \cite{Masanes2025responseto}, which was submitted to
Quantum along with the present paper; the papers were considered jointly and
both accepted.  

To preserve the historical flow of the discussion I have not altered the 
manuscript aside from this section, other than to correct one small typo, 
update references and add acknowledgements.
With the Editors' permission, I add a few comments on the debate here. 

There seem to be two main issues.   The first is the extent to which
Ref. \cite{masanes2019measurement} made tacit assumptions about the
class of theories under discussion and their properties, which may
have appeared natural from their theoretical perspective but needed
to be made explicit to derive their theorems.   
Stacey has also discussed this both prior to \cite{stacey2023masanesgalleymullerstateupdatepostulate} and responding to \cite{stacey2024contradictionscuriositieskentscritique} the present paper.
I agree with Ref. \cite{stacey2024contradictionscuriositieskentscritique} that
the original presentation was deficient: readers can, of course, make up their own minds.

Of course, authors may consider whatever postulates they wish, and there is
no doubt that the measurement postulates of quantum mechanics can be derived from
a sufficiently restrictive set of assumptions.  
The second (much more scientifically interesting) issue, then, is whether there are 
compelling reasons to make assumptions that exclude the types of counterexample
discussed in the present paper and other alternatives to the standard quantum
measurement postulates.    For the counterexamples discussed here, the answer
seems to turn on whether one regards the quantum no-signalling principle
(and not only the relativistic no-superluminal signalling principle) as 
sacrosanct. 

One argument for the quantum no-signalling principle is that, without it, 
we might be unable to do science.   Given that we can never access the 
quantum state of the whole universe, we need to make inferences about 
its structure and physical laws from observations on subsystems.
Violating quantum no-signalling allows the possibility that the
physics of any subsystem is constantly changing as the result of 
actions on (what we currently think of) distinct subsystems, including
physically distant ones.    
A completely general theory of this type, without any limitation on the 
rules by which actions on one subsystem could affect another, might indeed
make science very difficult.   However, the examples discussed in this
paper show that violating the quantum no-signalling principle is 
consistent with retaining those features of quantum 
subsystem composition that allow us to infer scientific laws from
observations on subsystems.   
The various devices we consider {\it extend} the range of possible measurements on
quantum systems, and mean that we can obtain strictly {\it more} information about
them.   All the scientific methodologies and tests that gave rise to and
confirm standard quantum theory remain available and valid.   Moreover, we
can test and confirm standard quantum theory even more precisely and compellingly,
by obtaining direct information about the quantum state of any system
-- and, in the examples presented, doing so non-destructively.
We can also test and confirm the properties of the devices, including the
effects on device readings on one subsystem caused by actions on other subsystems.
Whether or not this is a plausible hypothetical world, it is certainly one 
intelligible by standard scientific methods.   

Another argument is that violating the quantum no-signalling principle requires
some sort of unphysical magic, in which information becomes available in one
space-time region as a result of operations in a disjoint region, without any
physical carrier transmitting it.   We already discussed this above, but 
perhaps it is worth adding a few comments here.
The intuition that information transfer requires carriers is certainly
understandable and could be right.   
It is worth asking, though, how compelling we should find this sort of 
intuition.  After all, Einstein-Podolsky-Rosen's intuitions \cite{EPR35} about the properties of elements of physical reality were similarly understandable -- and indeed 
one might feel some intuitive tension between information-transfer-requires-carriers
and Bell nonlocality, although they are logically consistent.   
Even if one takes the intuition as dogma, it does not remove much of the scientific
motivation for considering our examples.   Our hypothetical devices define
extensions of quantum theory, but the extended theory need not be a complete
fundamental theory.   The underlying fundamental theory may include the
hypothesised information carriers.    For example, if some version of semi-classical gravity
holds in some regime, then, in that regime, measuring the classical gravitational field 
of a quantum system defines a type of state readout device \cite{kent2021quantum,fedida2024mixtureequivalenceprinciplespostquantum}.   
Since the gravitational field presumably exists everywhere, 
it (or classical variables associated with it) can be the hypothesised carrier(s).   

Given all this, I would argue that, 
while the quantum no-signalling principle may well indeed be a fundamental principle of nature, we need to rely on logic and experiment rather than
ruling out without further discussion the types of alternative measurements described above.  There is no hint that they are possible, and serious difficulties in making plausible relativistic theories that allow them.   If it can be shown that no 
such theory is possible, perhaps we should conclude that they are merely 
pathological counter-examples.   At present, though, they remain available as
theoretical possibilities, and testing them as foils for standard 
quantum theory (see e.g. \cite{kent2021quantum}) seems worthwhile.  

.

\section{Acknowledgements}
I thank Llu\'is Masanes, Thomas Galley, Markus M\"uller and Blake Stacey for
enjoyable, collegial and helpful discussions.   
I gratefully acknowledge financial support from UK Quantum
Communications Hub grant no. EP/T001011/1 and UK-Canada Quantum for Science research collaboration grant OPP640.  
This work was also supported by an FQXi grant and by Perimeter Institute
for Theoretical Physics. Research at Perimeter Institute is supported
by the Government of Canada through Industry Canada and by the
Province of Ontario through the Ministry of Research and Innovation.



\section*{References}
\bibliographystyle{plainnat}
\bibliography{qualia4}{}
\end{document}